\documentclass[prb,showpacs,aps,twocolumn]{revtex4}
\usepackage{amsmath}
\usepackage{graphicx}
\usepackage{dcolumn}
\usepackage{amssymb}

\begin{document}

\title{Impurity effects on semiconductor quantum bits in coupled quantum dots}
\author{Nga T. T. Nguyen and S. Das Sarma}
\affiliation{Condensed Matter Theory Center, Department of Physics,
University of Maryland, College Park, Maryland 20742-4111, USA}

\begin{abstract}
We theoretically consider the effects of having unintentional
charged impurities in laterally coupled two-dimensional double
(GaAs) quantum dot systems, where each dot contains one or two
electrons and a single charged impurity. Using molecular orbital and
configuration interaction method, we calculate the effect of the
impurity on the 2-electron energy spectrum of each individual dot as
well as on the spectrum of the coupled-double-dot 2-electron system.
We find that the singlet-triplet exchange splitting between the two
lowest energy states, both for the individual dots and the coupled
dot system, depends sensitively on the location of the impurity and
its coupling strength (i.e. the effective charge). A strong
electron-impurity coupling breaks down equality of the two
doubly-occupied singlets in the left and the right dot leading to a
mixing between different spin singlets. As a result, the maximally
entangled qubit states are no longer fully obtained in zero magnetic
field case. Moreover, a repulsive impurity results in a
triplet-singlet transition as the impurity effective charge
increases or/and the impurity position changes. We comment on the
impurity effect in spin qubit operations in the double dot system
based on our numerical results.
\end{abstract}

 \maketitle

\section{Introduction}

The goal of this work is the calculation of the low-lying energy
spectra of 2-electron semiconductor quantum dots (QDs) (both single
dots and coupled double-dots) in the presence of nearby static
charged impurity centers within a minimal model.  The purpose is to
quantify the  effects of quenched random charged impurities on the
singlet-triplet splitting in QDs in order to assess the importance
of background unintentional impurities, which are invariably present
in the environment, in adversely affecting the operations of
exchange-coupled dots as elementary spin qubits for solid state
quantum computation.  Since the positions and the strengths of the
background unintentional impurities are unknown and random, we study
the impurity effects as functions of impurity position and coupling
strength (defined as the effective impurity charge $Z$) assuming the
impurities to be Coulombic charge centers so that their effective
potential falls off slowly as a 1/r potential away from the impurity
location where `r' is the distance from the impurity.  Since the
main background impurities in GaAs and Si, the two most relevant
semiconductors of interest for solid state spin quantum computation,
are random charge centers, our consideration of Coulombic impurities
with a long-range impurity potential is reasonable.  The theoretical
results presented herein, while being completely microscopic and
fully quantum mechanical, are phenomenological in nature since the
impurity position and charge are treated as unknown parameters.
While our results show clearly the strong effect of local background
charged impurities on the low-lying QD spectra, its usefulness is
limited in comparing with experiment since no direct information
about impurity locations is currently available experimentally.  On
the other hand, our results establish the manifest importance of
using the highest quality background materials for semiconductor
spin qubit operations since the presence of even a single charged
impurity in the vicinity of the QDs seems to completely ruin the
operational logistics of  coupled QD systems.  The exchange coupling
(or equivalently, the singlet-triplet splitting) depends strongly
and sensitively on the location and the strength of the charged
impurity, which means that (1) a single charged impurity located
nearby could destroy the qubit, and (2) even a remote charged
impurity could have a strong adverse effect inducing substantial
qubit decoherence if the impurity is dynamic and has a fluctuation
timescale comparable to gate operations timescales-- in fact, the
impurity fluctuation timescale will become a dominant decoherence
time since the exchange energy will vary substantially over this
timescale.  The motivation of our work is a clear understanding of
the energetics of QD systems in the few electron situation in the
presence of charged impurities so that some rudimentary quantitative
magnitudes of the impurity effects are available for qubit operation
considerations.

Coupled QDs for quantum
computation\cite{Bennett,Kane,Loss,Vrijen,Hanson1} have been
extensively studied due to the prospect of using QDs as ideal
environment to confine and manipulate the QD electron spins. The
quantum bit, or qubit, of information is encoded and stored in these
localized single electron spins which exploit a spin relaxation time
of the order of milliseconds\cite{Elzerman2,Rugar,Petta3},
sufficiently long to allow the performance of coherent spin
operations. The proposed quantum computer\cite{Loss} in solid states
operates based on the exchange coupling $J$ between the two electron
spin qubits manipulated by an external magnetic field. This exchange
energy can be envisioned as the effective coupling between the two
spins in the double dots\cite{Loss} via the Heisenberg spin
Hamiltonian, $\widehat{H}=J \textbf{s}_1\cdot\textbf{s}_2$, which
takes into account possible contributions from different
hybridzation between singlet and triplet states. Hence, $J$ is
determined through the gate voltage control over the tunneling
coupling between the coupled two QDs. A complete understanding of
$J$ is important because it directly determines the $\sqrt{SWAP}$
operator which describes the exchange information between the two
qubits in the double dots. The fast solid-state two-qubit
operation\cite{Loss,Hu3}, generated as a consequence of the electron
spin exchange under electrical control, and the single-qubit
operation\cite{Koppen} suffice to assemble a standard quantum
computing system. The number of electrons in such a QD system can be
controlled precisely\cite{Ciorga,Elzerman,Poiro,Petta2} to 0, and
the electron exchange interaction is tunable by the applied gate
voltages. Thereby, the coupling between the dots can also be
controlled.

Established quantum regimes such as quantum entanglement between
individual electrons in one dot with the other electrons in the
other dots and superposition of electron spin qubits are the major
objects\cite{Loss,Vrijen,Tittel,Weihs,Burkard,Loss2,Rowe,Hu2,Burkard2,Hu,Hu3,Sausa,Levy,Harju}
in such exchange coupled QD system. Prerequisite criteria for
realization of a quantum computation system\cite{Loss} (including
initialization, manipulation of spin qubits, and readout) have been
demonstrated for single dot\cite{Hanson} and coupled QD
system\cite{Petta}, provided the long relaxation time
\cite{Elzerman2,Petta2} of the spins, by using charge sensing and
fast spin-to-charge conversion techniques. One of the perceived
advantages of solid-state quantum computing is the scalability with
existing semiconductors. In addition, integrating multi coupled dots
into a quantum circuit is made possible by adding more suitable gate
electrodes.\cite{Elzerman,Hanson1}

QDs in versatile GaAs semiconductors not only have been considered
the most widely-studied objects in the QD science but also their
well-understood physics are applicable to a variety of
materials.\cite{Hanson1} Often, unintentional impurities found in
the dot sample are used in the fabrication process to adjust the
potential well height between different heterostructure
semiconductors (such as Si in gated GaAs/AlGaAs QDs - see e.g.
Ref.\cite{Hanson1}) allowing the charge flow of electrons.
Statistically, such QD system containing impurities can be studied
for one, two, ... impurities. On the other hand, when integrating
multiple coupled 2-dot systems utilized as a multi-qubit gate, in
the non-overlapping regime between different 2-dot systems, a single
spin qubit in one coupled 2-dot system can be affected\cite{Weperen}
by the other coupled 2-dot systems. Each coupled 2-dot system acts
as a source of electrostatic field to the others and can be treated
as ``charged impurities".

Theoretical studies of impurities in coupled QDs have been rarely
found in the literature. In fact, the impurities are practically
found randomly in/outside the dot sample and their positions cannot
be specified precisely. In a coupled triple-dot system, a relatively
large collection of impurities was considered\cite{Irene}. These
statistical impurities were theoretically assumed to induce a weak
perturbation to the coupled triple QDs. The authors\cite{Irene}
found that the Coulomb exchange energy of the impurities with the QD
electrons in this study resulted in decoherence of the coded qubit
states. Thus, any information processing using electron charge
degree of freedom needs to take into account the decoherence channel
due to charge fluctuations.

In the present paper, we study the influence of charged impurities
on the singlet-triplet splitting of the two lowest energy levels as
well as the energy spectrum as a function of the impurity position,
the impurity effective charge, and the number of impurities for a
coupled two-dot system in zero magnetic field. We also examine the
impurity effect on the coupled electron qubits by tuning the
potential well height to different values and obtain different
triplet-singlet transitions for the repulsive impurity case. To
accomplish, we discuss the singlet-triplet splitting in the presence
of two impurities with similar charge located in the two separate
dots of the system.

The paper is organized as follows. In Sec. II we introduce the model
and methodology. Impurity effect on energy spectrum of a
two-electron single QD containing a single charged impurity is first
discussed in detail in Sec. III. Sect. IV is spent to present our
studies on a coupled 2-dot system in the presence of one or two
charged impurities. We examine the impurity position and impurity
effective charge dependence of the singlet-triplet spin splitting
energy. Influence of the confining potential barrier height on the
electron-impurity (e-I) and electron-electron (e-e) coupling is also
explored. All the results presented in this paper are obtained at
zero magnetic field, $B$=0. Summary of our results and conclusion
are found in Sec. V.

\section{Theoretical model}
Horizontally-coupled QDs are grown by the depletion of the
two-dimensional electron gas (2DEG) using typically the gated
mechanism\cite{Wiel,Petta0}. Such gated QDs have typical size of
about few tens of nm. Consequently, the lowest excitation energy in
such a QD system is found of the order of few meV. When the
inter-dot coupling strength is substantial, the electrons in the
coupled dots strongly quantum mechanically couple with each other.
As long as the phase is coherent, the electrons can ``tunnel"
between the adjacent dots forming different entangled qubits. Using
single-conduction-band effective-mass approximation which was
justified in Ref.\cite{Hu}, the Hamiltonian describing a
two-electron coupled double QD system containing unintentional
charged impurities (charge size $Ze$) in zero magnetic field reads:
\begin{equation}\label{e:Hamiltonian_D}
\widehat{H}=\widehat{h}_1+\widehat{h}_2+
\widehat{V}_C+\widehat{V}_{\mbox{e-Imp}}+\widehat{V}_{\mbox{I-I}}.
\end{equation}
Here, the first two terms $\widehat{h}_{i,j}$ are the
single-particle Hamiltonian of the two quantum-dot electrons
(coordinates $\textbf{r}_i$)
\begin{equation}
\widehat{h}_i=\frac{(i\hbar\bf{\nabla}_i+e\bf{A}_i)^2}{2m^*}+V(\bf{r}_i)
\end{equation}
confined by a potential well $V(\textbf{r})$. In the present paper,
this confining potential is constructed as a linear combination of
the three different Gaussians and can be separated into two parts:
\begin{eqnarray}\label{e:confining_D}
V(\bf{r})&=&V_0\Large{(}e^{-[\frac{(x-a)^2}{l_x^2}+\frac{y^2}{l_y^2}]}+e^{-[\frac{(x+a)^2}{l_x^2}+\frac{y^2}{l_y^2}]}\Large{)}\\\nonumber
&&+V_b e^{-(\frac{x^2}{l_{bx}^2}+\frac{y^2}{l_{by}^2})}.
\end{eqnarray}
The first part acts as a double-well confining potential for the
coupled double dot system and the second part is used to control the
electrostatic barrier height between the two dots independently. The
set of varying parameters $V_0, V_b, l_x, l_y, l_{bx}, l_{by}$
characterizes the potential well depth and barrier height. A
nonvanishing overlap between the wave functions of the two quantum
dots signifies the electron virtual tunneling between the two dots,
i.e. the exchange energy is nonzero. Because this property can be
tuned by the applied gate voltage this exchange energy thereby can
be controlled. It is worth noting that $V_b$ can independently
modulate the barrier height resulting in concomitant change in the
inter-dot separation without modifying the single-particle energy
spectrum of the individual QDs in the system. %%
\begin{figure}[hbtp]
\begin{center}
\vspace*{-0.5cm}\leftskip-0.0cm
\includegraphics[width=8.5cm]{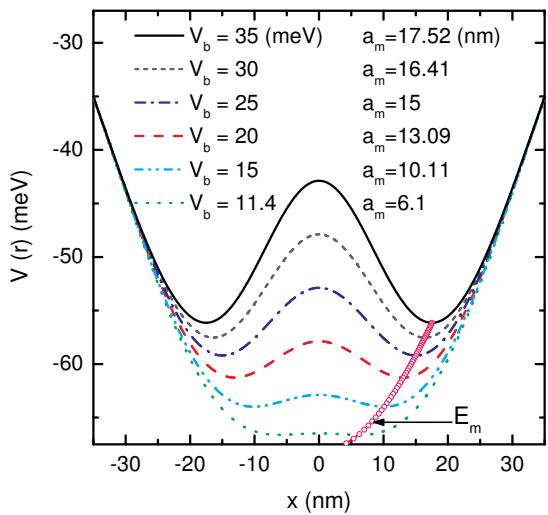}
\end{center}
\vspace{-0.7cm} \caption{(Color online) Confinement potential double
well of coupled double QDs modeled along the $x$-direction plotted
for different barrier depth $V_b$ (from 35 down to 11.4 meV) (left
column). The corresponding double well minima $(\pm{a_m},0)$ are
indicated in the same line in the right column of the reference. The
confinement energy at the two well bottoms $E_m$ is enclosed as the
magenta open circles for various values of $V_b$.}
\label{double-dot_potential}
\end{figure}
%%
%%%Them tu report_1
The last three terms in Eq.~(\ref{e:Hamiltonian_D}) are the Coulomb
interaction between, respectively, the two electrons
\begin{equation}
\widehat{V}_C=\frac{e^2}{4\pi\epsilon\epsilon_0|\bf{r}_{ij}|},
\end{equation}
the electron and the impurities
\begin{equation}
V_{\mbox{e-Imp}}=\sum_{k=1}^{N_I=2}\sum_{i=1}^{N_e=2}\frac{Z_ke^2}{4\pi\epsilon\epsilon_0|\bf{R}_k-\bf{r}_i|}
\end{equation}
where $N_I$ and $N_e$ denote the number of impurities and electrons,
respectively, and the impurities
\begin{equation}
V_{\mbox{I-I}}=\frac{Z_1Z_2}{4\pi\epsilon\epsilon_0|\bf{R}_1-\bf{R}_2|}.
\end{equation}

The solution of a single particle confined by a parabolic potential
is well known as the Fock-Darwin basis:
\begin{equation}\label{e:Fock-Darwin}
\varphi_{nl}\left(\bf{r}\right)=
\frac{1}{l_0}\sqrt{\frac{n!}{\pi\left(n+|l|\right)!}}\left(\frac{r}{l_0}\right)^{|l|}
e^{-il\theta}e^{-\frac{r^2}{{2l_0^2}}}L_n^{|l|}\left(\frac{r^2}{l_0^2}\right)
\end{equation}
with corresponding energy
\begin{equation}
E_{n,l}=\hbar\omega_0(2n+|l|+1)
\end{equation}
where $(n,l)$ stand for radial and azimuthal quantum number,
respectively. $l_0$ is the confinement length which is defined via
the confinement frequency $\omega_0$:
$l_0=\sqrt{\frac{\hbar}{m_e^*\omega_0}}$. Using this basis as the
radial part we can construct the many-body wave function for the
considered
system.  %It is
%important to note that the number of single-particle states is taken
%sufficiently large to guarantee the numerical accuracy.

We introduce a dimensionless parameter,
\begin{equation}\label{alpha}
\lambda=\frac{l_0}{a_B^*}
\end{equation}
with $a^*_B=\frac{4\pi\epsilon_0\epsilon\hbar^2}{m_e^*e^2}$ the
effective Bohr radius, which is used to describe the relation
between the effective Rydberg energy $R_y^*=\frac{m^*
e^4}{2\hbar^2({4\pi\epsilon_0\epsilon})^2}$ and the confinement
energy $\hbar\omega_0$
\begin{equation}
R_y^*=\hbar\omega_0\cdot\frac{\lambda^2}{2}.
\end{equation}
Both e-e and e-I Coulomb interaction are evaluated in terms of
matrix elements as
\begin{eqnarray}\label{Coulomb}
\langle\Psi|V_{\mbox{e-e}}|\Psi\rangle&=&V^C_0\langle\Psi|\frac{1}
{\bf{\widetilde{r}}_{i}-\bf{\widetilde{r}}_j|}|\Psi\rangle\\\nonumber
\langle\Psi|V_{\mbox{e-Imp}}|\Psi\rangle&=&V^C_0\langle\Psi|\frac{1}{|\bf{\widetilde{r}}-\bf{\widetilde{R}}|}|\Psi\rangle
\end{eqnarray}
where $\Psi$ denotes the wave function of the system,
$\widetilde{\textbf{r}}_{i,j}=\textbf{r}/l_0$,
${\widetilde{\textbf{R}}}=\textbf{R}/l_0$, and
$V_0^C=e^2/4\pi\epsilon\epsilon_0l_0$ is the Coulomb energy unit.
$V_0^C$ relates to the confinement energy via the relation:
\begin{equation}
V_0^C=\frac{e^2}{4\pi\epsilon\epsilon_0l_0}=\lambda \cdot
\hbar\omega_0.
\end{equation}

The numerical results are implemented for e.g. a GaAs QD with
$m^*=0.067m_0$, $\epsilon=13.1$, $g_e=-0.44$, $R_y^*=5.31$ meV
(corresponding to $a_B^*=10.3$ nm). $\lambda$ is changed by changing
the confinement energy $\hbar\omega_0$. $\lambda=1$ gives
$\hbar\omega_0=2R_y^*=10.62$ meV and corresponding confinement
length $l_0=a_B^*=10.3$ nm.

For Si/SiGe ($\epsilon$=13) and Si/SiO$_2$ ($\epsilon$=6.8) quantum
dots with a heavier effective mass $m^*=0.19m_0$, effective Bohr
radius and effective Rydberg energy are, respectively, $a_B^*=4$ nm
and $R_y^*=15$ meV, and $a_B^*=2.11$ nm and $R_y^*=44.76$ meV (see
e.g. Ref.\cite{Qiuzi}). $\lambda=1$ gives $\hbar\omega_0=2R^*_y=30$
meV ($l_0=4$ nm) for a Si/SiGe QD and
$\hbar\omega_0=2R^*_y\approx90$ meV ($l_0=2.11$ nm) for a Si/SiO$_2$
QD system. Even though our numerical results are applicable to GaAs
QD systems, singlet-triplet splitting energies for Si QD systems are
also provided in Appendix~\ref{A:Si} for comparison purposes.

Reducing $\lambda$ means that the effective length, $l_0$, of the QD
decreases while the energy spacing, $\hbar\omega_0$, between the 2D
shells, i.e. the $s$-, $p$-, $d$- levels, will increase. In the
small $\lambda$ limit, the problem at hand converts to the problem
of independent particles. In the opposite case, very large
$\lambda$, the problem approaches the classical situation.

In the coupled double QD system, the single-particle solutions in
each dot are obtained approximately based on an assumption that
around the center of the each dot ($\pm a_m,0$) the single-electron
problem can be treated as a 2D harmonic oscillator. This means that
the confining potential well $V(\textbf{r})$ performs a quadratic
form $V(\textbf{r})-E_{m}\approx \frac{V_0}{l_x^2} [(x\pm
a_m)^2+y^2]$, with $E_{m}$ the bottom energy of the potential well,
around $(\pm{a_m},0)$. Changes in $E_m$ when $V_b$ varies can be
obtained as the open circles in Fig.~\ref{double-dot_potential} The
single-particle wave functions now are identical to the Fock-Darwin
levels centered at $(\pm{a_m},0)$ and the single-particle energy
spectrum is shifted by an amount of $E_m$.
\begin{eqnarray}\label{e:FDam}
\varphi_{L(R)}\left(\bf{r}\right)&=&
\frac{1}{l_0}\sqrt{\frac{n!}{\pi\left(n+|l|\right)!}}\left(\frac{r_{L(R)}}{l_0}\right)^{|l|}
e^{-il\theta}\\\nonumber&&\,\,\,\,\,\,e^{-\frac{r_{L(R)}^2}{{2l_0^2}}}
L_n^{|l|}\left(\frac{r_{L(R)}^2}{l_0^2}\right)
\end{eqnarray}
where $\textbf{r}_{L,R}=(x\pm a_m,y)$ and
$\omega_0=\sqrt{\frac{2V_0}{m^*l_x^2}}$ is the quadratic confining
frequency which defines a new length
$l_0^2=\frac{\hbar}{m^*\omega_0}$ called the confinement length. The
single-particle ground-state wave function is
\begin{equation}
\varphi_{L,R}=\frac{1}{\sqrt{\pi}l_0} e^{-\frac{[(x\pm
a_m)^2+y^2]}{2l_0^2}}.
\end{equation}
To quantitatively evaluate the advantage of using the above
confining potential present in Eq.~(\ref{e:confining_D}), we plot in
Fig.~\ref{double-dot_potential} the confining potential well where
its barrier height and the QD centers are modified by changing
$V_b$. Here, $V_b$ is reduced from $35$ meV to $30, 25, 20, 15$, and
$11.4$ meV. As a result, the barrier height will decrease making the
electron exchange energy increased. For example, the system with
$V_b=35$ meV has the corresponding $a_m\approx17.52$ nm ($\approx
1.75 l_0$), and barrier height $\Delta{V_b}=13.26$ meV. For
$V_b=30$meV, these parameters are $a_m\approx16.41$ nm ($\approx
1.64l_0$) and $\Delta{V_b}=9.65$ meV. Decreasing $V_b$ leads to a
shortened inter-dot separation $2a_m$ and a smaller $\Delta{V_b}$.
Details can be obtained in Fig.~\ref{double-dot_potential}. In our
numerical results, we use $V_b=30$ meV for most of our calculations
except when we examine the barrier height dependence of the exchange
energy $J$ where $V_b$ can vary. The center region of each QD,
however, has unchanged effective length, namely $l_0$, regardless to
the change in the barrier height. The other parameters taken after
Refs.\cite{Hu,Hu3,Sausa} are $l_x=l_y=30$ nm, $a=l_{bx}=l_x/2=15$
nm, $l_{by}=80$ nm.

%%dang viet
We assume that the impurities are located arbitrarily in or outside
the coupled QDs. Their coordinates are
$\textbf{R}_k=({x_k},{y_k},{z_k})$, \{k=1,2\}. Theoretically, the
effective coupling between electron and the localized impurities as
well as the coupling between the impurities with each other can be
tuned by engineering the impurity charge $Z$.

Configuration interaction (CI) and molecular orbital (MO) method are
used to numerically solve the Hamiltonian
Eq.~(\ref{e:Hamiltonian_D}). Both construct the total wave function
of the system as a superposition of different possible quantum
configurations (CI) or molecule states (MO) extended in the basis of
single-particle wave functions:
\begin{equation}
\Psi(\textbf{r}_1,\textbf{r}_2)=\sum_i^{N_c}\psi_i(\textbf{r}_1,\textbf{r}_2)
\end{equation}
where $\psi_i(\textbf{r}_1,\textbf{r}_2)$ represents one
many-electron configuration as a Slater determinant. Each term of
this Slater determinant is a single-electron wave function consisted
of the radial part, the Fock-Darwin state
$\varphi$($\textbf{r}_1$,$\textbf{r}_2$) defined in
Eq.~(\ref{e:FDam}), and the electron spin part (detailed
justifications are referred to Ref.\cite{Hu}).

The singlet-triplet spin splitting energy of the electrons $J$ in
such a coupled 2-dot system is defined as the energy difference
between the two lowest singlet ($\Psi^S$) and triplet ($\Psi^T$)
states:
\begin{equation}
J=\langle{\Psi^T}|\widehat{H}|\Psi^T\rangle-\langle{\Psi^S}|\widehat{H}|\Psi^S\rangle.
\end{equation}

%In this section, we will describe the model of coupled two-electron
%QDs in the presence of a single or two impurities. The two
%electrons are confined by a Gaussian potential for a coupled dot
%system, whose form is taken from the confinement potential in
%Xuedong Hu and Das Sarma's paper, and both interact with the
%impurity through a Coulomb interaction. The strength of the
%e-I interaction can be tuned by changing the impurity
%effective charge or its position in/outside the QD system.
%We will also discuss the employed approaches which are the
%Hund-Mulliken and configuration interaction method.
\subsection{Hund-Mulliken}
In Hund-Mulliken model, the energy spectrum consists of four levels
which are four possible superpositions of the four basis (three
singlets and one triplet) wave functions
\begin{eqnarray}\label{e:Bell}
\psi_1^S(\textbf{r}_1,\textbf{r}_2)&=&\frac{1}{\sqrt{2}}[\varphi_L(\textbf{r}_1)\varphi_R(\textbf{r}_2)+\varphi_L(\textbf{r}_1)\varphi_R(\textbf{r}_2)]\\\nonumber
\psi_2^S(\textbf{r}_1,\textbf{r}_2)&=&\varphi_L(\textbf{r}_1)\varphi_L(\textbf{r}_2)\\\nonumber
\psi_3^S(\textbf{r}_1,\textbf{r}_2)&=&\varphi_R(\textbf{r}_1)\varphi_R(\textbf{r}_2)\\\nonumber
\psi^T(\textbf{r}_1,\textbf{r}_2)&=&\frac{1}{\sqrt{2}}[\varphi_L(\textbf{r}_1)\varphi_R(\textbf{r}_2)-\varphi_L(\textbf{r}_1)\varphi_R(\textbf{r}_2)].
\end{eqnarray}
In the coupled QD system without an impurity, these three singlets
do not couple with the maximally entangled triplet state $\psi^T$
therefore they can be treated separately. The entire Hamiltonian
matrix represents these singlets and triplets as independent blocks.

\subsection{Configuration Interaction}

The most difficult task in finding the eigenvalues of a coupled
2-dot system lies in the basis choice among which Fock-Darwin basis
and Gaussian basis are the most widely used. However, in both cases
a closed analytical form for the Hamiltonian matrix elements,
essentially the e-e Coulomb matrix elements, has not yet been
obtained. The reason is that the single-particle solutions of
different dots i) have distinct zero-points shifted to the two
bottoms of the confining potential well and ii) thus are not
orthogonal with each other. Consequently, the number of
distinguishable single-particle quantum states for the double-dot
system will be doubled. Effectively, the size of the entire
Hamiltonian increases in comparison to the single dot case.
Quantitatively, for $S_z=0$ subspace, such a number is 4 times
larger than that of the single dot problem (2NX2N). Specifically, if
only the $s$-waves are taken into account, the number of
configurations in the subspace $S_z=0$ is 1 for single QD and 4 for
coupled 2-dot system and if the $s$- and $p$-waves are included
those numbers are 9 and 36, respectively.

The above fact poses much difficulty to solving the eigenvalues of
the coupled QD Hamiltonian. The most time consuming part is spent to
calculate the Coulomb matrix elements. In the problem at hand, the
number of Coulomb elements increases due to the e-I exchange
interaction. However, based on the following facts one can take out
nonphysical excited single-particle quantum states: i) the form of
the confining potential, which is tuned electrostatically by metal
top gates, is not exactly known and ii) the barrier potential height
is finite. The latter factor validates the harmonic approximation
for the confinement potential double well resulting in only a
limited number of Fock-Darwin states involved.

\section{Impurity effects in single quantum dots}
In the single QD case, we assume that there is only one impurity and
the impurity is located along $z$-axis, i.e. $R=(0,0,d)$. The
addition Coulomb interaction of the electrons with the impurity is
obtained analytically (see Appendix~\ref{A:Coulomb}):
\begin{widetext}
\begin{eqnarray}\label{V_matrix}
%V^{n_2l_2}_{n_1l_1}(R)=\delta_{l1,l2}A^{l^+}_{n_1n_2}\sum_{j=0}^{n_1}\sum_{k=0}^{n_2}B_{jk}I_{m}(R)\\
V^{n_2l_2}_{n_1l_1}(\widetilde{R})&=&\delta_{l1,l2}\frac{2}{\sqrt{\pi}}\frac{\Gamma\left(n_1+l^{+}+1\right)\Gamma\left(n_2+l^{+}+1\right)}{\sqrt{n_1!n_2!(n_1+l^{+})!(n_2+l^{+}+1)!}}\times\\\nonumber
&&\sum^{n_1}_{j=0}\sum^{n_2}_{k=0}\frac{(-n_1)_{j}(-n_2)_{k}\Gamma(l^{+}+j+k+1)\Gamma(l^{+}+1)}{j!k!\Gamma(l^{+}+j+1)\Gamma(l^{+}+k+1)}I_{m}(\widetilde{R})
\end{eqnarray}
\end{widetext}
where $l^{+}=|l_1|=|l_2|$ and
\begin{equation}\label{Im}
I_m(\widetilde{R})=\int_{0}^{\infty}\frac{e^{-{\widetilde{R}}^2u^2}}{(1+u^2)^{m+1}}du.
\end{equation}
Integral (\ref{Im}) can be obtained through the recurrence:
%\begin{equation}
%I_0(\widetilde{R})=\frac{\pi}{2}e^{\widetilde{R}^2}\mbox{erfc}(\widetilde{R}),
%\end{equation}
\begin{eqnarray}
I_0(\widetilde{R})&=&\frac{\pi}{2}e^{\widetilde{R}^2}\mbox{erfc}(\widetilde{R}),\\\nonumber
I_1(\widetilde{R})&=&\frac{1}{2}[(1-2\widetilde{R}^2)I_0{\widetilde{R}}+\sqrt{\pi}\widetilde{R}],\\\nonumber
I_m(\widetilde{R})&=&(-\frac{\widetilde{R}^2}{m}+1-\frac{1}{2m})I_{m-1}(\widetilde{R})+\frac{\widetilde{R}^2}{m}I_{m-2}(\widetilde{R})
\end{eqnarray}
where erfc$(x)$ is the complementary error function,
\begin{equation}
\mbox{erfc}(x)=\int_{x}^{\infty}\frac{e^{-x^2t^2}}{1+t^2}dt,
\end{equation}
which rapidly decreases in $x$. $(x)_n=x(x+1)(x+2)...(x+n-1)$ is the
Pochhammer function. Several values of the complementary error
function which can be obtained in any numerical library are:
erfc$(0)=1$, erfc$(0.01)\approx0.98872$, erfc$(0.1)\approx0.88754$,
erfc$(0.5)\approx0.47950$, erfc(1)$\approx0.1573$,
erfc(2)$\approx0.00468$, etc.

\subsection{Perturbative and exact calculations}
Following the e-I interaction whose matrix elements are expressed in
Eq. (\ref{V_matrix}), there is mixture between different states with
different total radial quantum numbers. Consequently, the
ground-state wave function e.g. for $N_e=1$ electron is no longer
the only one $s$- ($n$=0,$l$=0) quantum state. Instead, it consists
of several quantum states whose contributions $C_i$ are different
\begin{equation}
\Psi=\sum_{i=1}C_i\psi_i.
\end{equation}
For $N_e=2$ electrons without impurity, by increasing $\lambda$
results in the ground state mixing of different configurations. The
two electrons will occupy higher Fock-Darwin states with decreasing
energy spacing to lower their Coulomb repulsion. The easiest way to
check the accuracy of the numerical results when the impurity is
present is to consider the limitation when weak perturbation works
perfectly. The first-order perturbation approximation:
\begin{equation}\label{E}
E=E_0+Z\sum_{i=1}^{N_e}<\Psi_0|V(|\textbf{r}_i-\textbf{R}|)|\Psi_0>+O(Z^2)
\end{equation}
where $E_0$ is the GS energy of the QD with the ground-state wave
function $\Psi_0$ in case without impurity is a good one when $Z<1$.
For the sake of simplicity in evaluating the perturbative part in
Eq. (\ref{E}) for both $N_e=1$ and $2$ electrons, we assume that
$\Psi_0$ in case $Z=0$ is a single term: the $s$-state. For $N_e=1$
electron, this is always satisfied and its radial part is
\begin{equation}\label{s-state}
\Psi_0(r)=\frac{1}{l_0\sqrt{\pi}}e^{-\frac{r^2}{2l_0^2}}.
\end{equation}
\begin{figure}[hbtp]
\begin{center}
\vspace*{-0.5cm}\leftskip-0.3cm
\includegraphics[width=8.8cm]{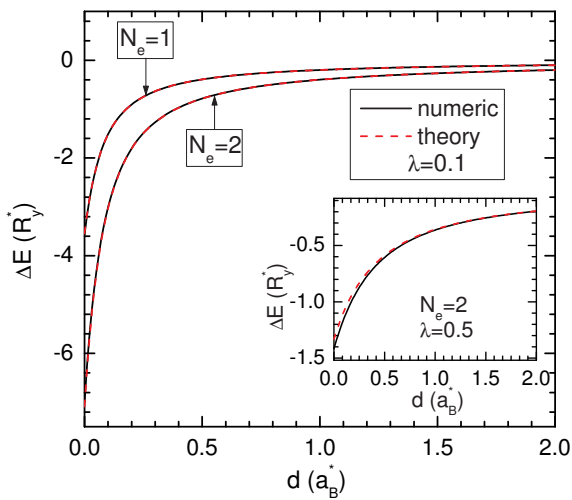}
\end{center}
\vspace{-0.8cm} \caption{(Color online) Agreement between
theoretically perturbative (red dash-dotted) and exact
diagonalization (black solid) calculations in the shift of the
ground-state energy due to the presence of the impurity in
single-electron (two upper curves) and two-electron (two lower
curves) QDs for $\lambda=0.1$. The impurity effective charge is
$Z=-0.1$. The inset shows addition data for the case that $\lambda$
is increased to 0.5 to examine whether the perturbative calculations
hold reliable. This energy shift is identical to the binding energy
of the system and scaled to the Rydberg energy $R_y^*=5.31$ meV for
GaAs QDs. } \label{1etheory}
\end{figure}

The situation changes for $N_e=2$. The above condition for $\Psi_0$
to be a single term is only satisfied when $\lambda<1$. If so, the
two electrons obeying Pauli exclusion principle with spins
antiparallel will mainly stay in the $s$-orbital in case no impurity
is present because their Coulomb repulsion is small as compared to
the confining energy. The ground-state wave function is
\begin{equation}
|\Psi_0>=c^{\dagger}_{s\uparrow}c^{\dagger}_{s\downarrow}|0>.
\end{equation}

Using this assumption, the total energy is estimated theoretically
for the single-electron QD:
\begin{equation}\label{E-r1}
E_{(N_e=1)}=E_{0_{(N_e=1)}}+2{\sqrt{\pi}}Z\lambda
\hbar\omega_0e^{\widetilde{R}^2}\mbox{erfc}\left({\widetilde{R}}\right)+O(Z^2)
\end{equation}
and the two-electron QD:
\begin{equation}\label{E-r2}
E_{(N_e=2)}=E_{0_{(N_e=2)}}+4{\sqrt{\pi}}Z\lambda
\hbar\omega_0e^{\widetilde{R}^2}\mbox{erfc}\left({\widetilde{R}}\right)+O(Z^2).
\end{equation}

We plotted in Fig.~\ref{1etheory} the energy shift due to the
presence of the impurity as a function of $d$ using the above
theoretical estimations (red dash-dotted) and the numerical results
(black solid) for $Z=0.1$ and $\lambda=0.1$ ($l_0=1.03$nm and
$\hbar\omega_0=200 R_y^*=1.062$eV) for $N_e=1$ (upper curves) and
$2$ (lower curves) electrons. Both theoretical and numerical results
are in good agreement.

A small comment is made in case $d$ is large. As seen from
Fig.~\ref{1etheory}, the e-I interaction goes to zero slowly when
$d$ increases to a relatively large value, say $d>a_B^{*}$. The
answer lies in the product
$\Big{[}e^{\widetilde{R}^2}\cdot$erfc$\left(\widetilde{R}\right)\Big{]}$
where the exponential function increases in $d$ competitively with
the complementary error function which decreases in $d$.

The first-order perturbation theory works very well as long as $Z<1$
and $\lambda$ is small enough. Because if $\lambda$ increases
different e-I couplings of different configurations will occur which
result in the presence of a substantial number of nonzero
off-diagonal terms. This fact leads to an increasing difference
between the first-order calculations and the numerical results. For
example, in the two-electron QD as plotted in the inset of
Fig.~\ref{1etheory}, this difference for $\widetilde{R}=0$, which is
also the largest value, is about $6\%$ when $\lambda$ increases to
$0.5$. This $6\%$ come from the configurations with minor
contributions ($n_1=1,l_1=0;n_2=1,l_2=0$), (0,1;0,-1), etc.
$n_1,l_1$ and $n_2,l_2$ are the radial and azimuthal quantum numbers
of the two electrons.

Practically, one often has GaAs QDs with $\lambda>1$ i.e. larger QDs
need to be considered. To describe such QDs, numerical calculations
are used for different values of $\lambda$. Here,  we examine the
case of $\lambda=2$ which has $l_0=2a_B^{*}=20.6$ nm and the
corresponding $\hbar\omega_0=R_y^*/2=2.655$ meV. The Coulomb
interaction unit ($V_0^C=5.31$ meV) in such a QD system is ($2$
times) larger than the confining energy $\hbar\omega_0$.

%In this subsection, we will discuss the agreement between
%perturbation and configuration interaction method by showing the
%convergence of the results obtained by these two methods for single-
%and two-electron dots when the e-I interaction is
%weak. Fig. 1 in the report will be included here
%%
\begin{figure}[hbtp]
\begin{center}
\vspace*{-0.5cm}\leftskip-0.3cm
\includegraphics[width=9.6cm]{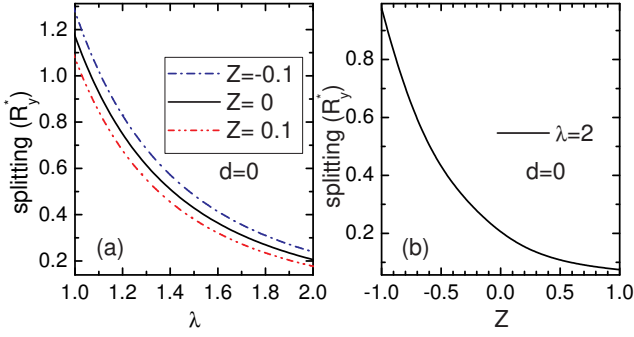}
\end{center}
\vspace{-0.8cm} \caption{Splitting energy of the two lowest singlet
and triplet energy levels as a function of (a) Coulomb interaction
strength $\lambda$ for three different $Z$=-0.1 (blue dashed-dot), 0
(black solid), and 0.1 (dash-dotted dot) and (b) impurity effective
charge $Z$ within the range (-1,1) for the case $\lambda=2$. The
impurity is located at the origin of the single two-electron QD:
d=0.} \label{single_Zchange_few_d}
\end{figure}
\subsection{Singlet-triplet splitting energy}
The Coulomb interaction between the QD electrons strongly competes
with the confining energy and with the e-I interaction as $Z$ and/or
$\lambda$ increase. Increasing $\lambda$ means that the confining
energy becomes smaller with respect to the Coulomb interaction. As a
result, electrons start to occupy higher Fock-Darwin levels. At zero
B-field and in the absence of impurity, the ground state consists of
several Fock-Darwin states where the dominant component is the
$s$-wave term. In the presence of a charged impurity, those electron
configurations that fulfill $L$=const form different $L$-subgroups
with nonzero contributions to the total wave function of the system.
Increasing the effective charge $Z$ results in a strong mixing
between those subgroups. Consequently, there are more than 1 state
which play as dominant components to the total wave function. We
plot in Figs.~\ref{single_Zchange_few_d}(a) and (b), respectively,
the splitting energy between the ground state (singlet) and the
first excited state (triplet) as a function of $\lambda$ and $Z$ in
case the impurity is located at the center of the single QD. As
$\lambda$ increases [see Fig.~\ref{single_Zchange_few_d}(a)], the
splitting becomes smaller for both with and without impurity cases.
It is because the singlet and the triplet states have many similar
nonzero configurations. Details are found below from the discussion
for particular values of $Z$. From
Fig.~\ref{single_Zchange_few_d}(a) we also notice that the energy
splitting shift due to the presence of the impurity remains almost
unchanged by changing $\lambda$ for both cases $Z=-0.1$ (blue dashed
dot) and $Z=0.1$ (red dash-dotted-dot). It is because the impurity
location is examined at the center of the QD $d$=0. Such splitting
shift will change if the impurity is displaced to any other
off-center position $d\ne0$.

The $Z$-dependence of the splitting energy in
Fig.~\ref{single_Zchange_few_d}(b) for $\lambda=2$ shows a
continuous decrease as $Z$ changes its sign from negative (-1,0) to
positive (0,1). The decrease is significant around $Z$=-1 which
coincides with the physics discussed above for the negative $Z$=-0.1
and -1 cases.

It is found that for $Z=-0.1$ and $-1$ the ground state and the
first-excited state in the two-electron QD containing a single
charged impurity are the singlet and the triplet, respectively.
Their major components are, respectively, the $s$-$s$ and the
$s$-$p$ configuration, i.e. one electron is in the s- and the other
in the $s$-($p$-)orbital. Note that the $s$-$s$ overlap has only one
maximum at the origin. In opposite, the $s$-$p$ overlap exhibits a
minimum at the origin.
\begin{figure}[hbtp]
\begin{center}
\vspace*{-0.5cm}\leftskip-0.6cm
\includegraphics[width=10.cm]{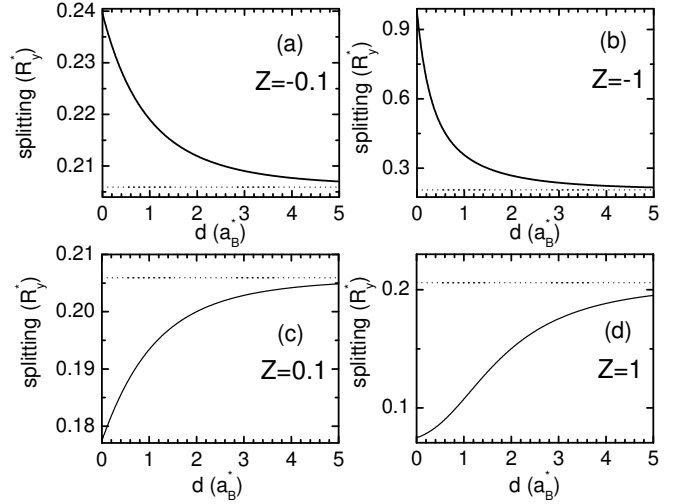}
\end{center}
\vspace{-1cm} \caption{Splitting of the two lowest singlet and
triplet states as a function of the impurity position in a
two-electron single QD for four different effective charges of the
impurity (a) $Z=-0.1$, (b) -1, (c) 0.1, and (d) 1 in case
$\lambda=2$ ($\hbar\omega_0=2R^*_y/4=2.655$ meV). Dotted lines are
the data obtained for the zero effective charge case ($Z=0$) for
reference. } \label{split}
\end{figure}
%%%

With changing the position of the impurity, the energy spin
splitting of the singlet-triplet states is obtained in
Fig.~\ref{split}(a) for $Z$=-0.1 and Fig.~\ref{split}(b) for $Z$=-1.
The presence of the impurity results in an increase in the splitting
which is largest when the impurity is located at the center of the
dot [as illustrated in Fig.~\ref{split}(a) and Fig.~\ref{split}(b)
for $Z=0$ (dot lines) and for $Z<0$ (solid lines)].

The magnitude of the splitting over an impurity charge unit
$\Big[\left(E^{I}_1-E^{I}_0\right)\equiv\mbox{splitting}\Big]$/Z,
with $E_{0,1}^{I}$ the energy of the ground state and the first
excited state in the doped QD, is found smaller for the larger $Z$
case [see Figs.~\ref{split}(a) and (b)] for a certain $d$ value.
This means that the e-I attraction becomes dominant over the e-e
Coulomb interaction. As a result, each state consists of different
configurations  with compatible contributions to the total wave
function. For example, by increasing the e-I effective strength from
$Z=-0.1$ [Fig.~\ref{split}(a)] to $Z=-1$ [Fig.~\ref{split}(b)] the
contribution of the singlet $s$-configuration decreases from
$C_0^2\approx0.83$ to respectively $C_0^2\approx0.64$. The
compensatory parts come from other configurations which also have
$L=0$ such as ($n_1=0,l_1=0;n_2=1,l_2=0$), (0,0;2,0), (0,0;3,0),
etc. Those states stay closer in energy with increasing $\lambda$
($=2$ in this case).

Above we discussed the impurity effect for an attractive impurity
(positively charged $Z<$0). A negatively charged impurity which
induces a repulsive coupling with electron is examined in
Figs.~\ref{split}(c) and (d) for two effective charges $Z$=0.1 and
Z=1. The splitting energy between the ground state singlet and the
first excited triplet is plotted as a function of the impurity
position $d$. Apparently, when the impurity is found at the origin
$d$=0, it repels the two electrons most. The probability of the
electrons to be found at the origin ($s$-orbital) reduces
significantly. In this case, the electrons can be found at other
higher Fock-Darwin states. This means that the ground state energy
becomes closer in energy to the first excited state. That explains a
smaller splitting energy we obtain for $Z$=0.1 [see
Fig.~\ref{split}(c)] and $Z$=1 [see Fig.~\ref{split}(d)] as compared
to, respectively, the cases $Z$=-0.1 [see Fig.~\ref{split}(a)] and
Z=-1 [see Fig.~\ref{split}(b)]. Moreover, we see from
Figs.~\ref{split}(c) and (d) that when the impurity is moved away
from the center of the QD, the splitting energy starts to increase.
In other words, the probability of finding the electrons in the
$s$-orbital increases.

\subsection{Impurity effect on the energy spectrum}

Technically, the presence of a charged impurity leads to an increase
in the number of nonzero off-diagonal elements of the Hamiltonian
matrix. We present such examination on the energy spectrum of the
two-electron single QD as a function of the impurity position. The
competition between the two types of Coulomb interaction results in
different relative orders of the energy levels depending on both the
impurity charge $Z$ and the impurity position $d$.
\begin{figure}[hbtp]
\begin{center}
\vspace*{-0.5cm}\leftskip-0.1cm
\includegraphics[width=8cm]{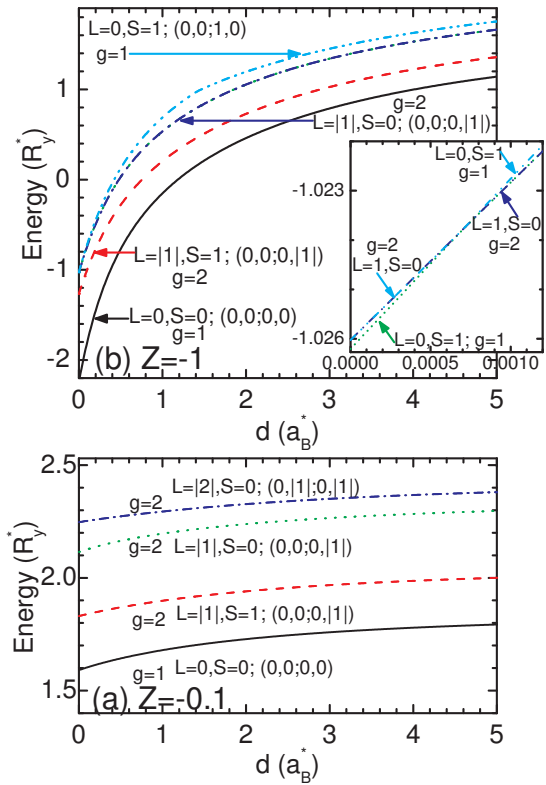}
\end{center}
\vspace{-1cm} \caption{Low-level energy spectrum as a function of
$d$ of a two-electron QD in case (a) $Z=-0.1$ and (b) $Z$=-1 for
$\lambda=2$. Each level is labeled by: g as level's degeneracy, $L$,
$S$ as total angular momentum and total spin, and the major electron
configuration ($n_1,l_1,n_2,l_2$). $n_1, l_1$ and $n_2, l_2$ are the
radial and azimuthal quantum numbers for electron 1 and 2. Inset in
(a) is a magnification of the main plot which highlights the region
with the occurrence of the crossing at $d/a^*_B\approx0.0008$. }
\label{specZ-0.1}
\end{figure}
%%%
%
\begin{figure}[hbtp]
\begin{center}
\vspace*{-0.5cm}\leftskip0.5cm
\includegraphics[width=8.3cm]{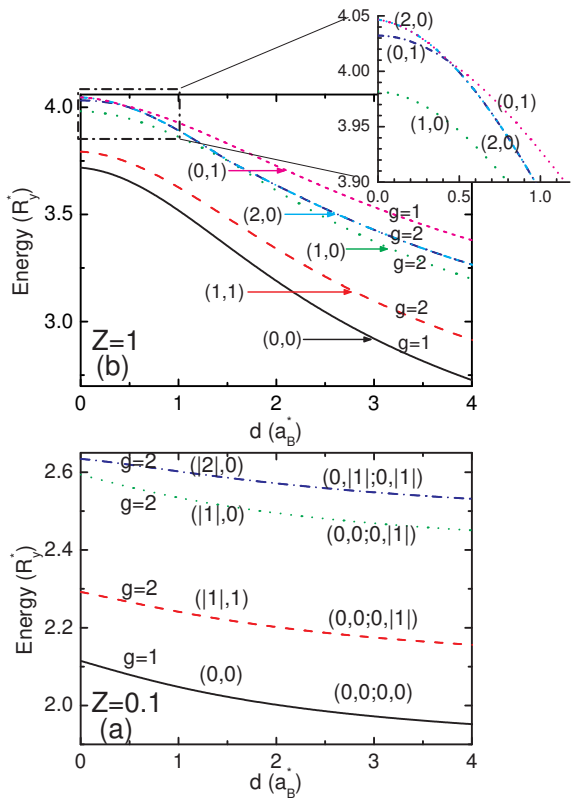}
\end{center}
\vspace{-1cm} \caption{Low-level energy spectrum of two-electron
single QD containing a repulsive impurity (a) $Z$=$0.1$ and (b) $1$
for $\lambda=2$. Inset in (a) is the magnification of the main plot
into the region $d$=($0,1.1a^*_B$) to signify the crossing at
$d=0.44a^*_B$ between the states (L=0,S=1) and (L=2,S=0).}
\label{specPositive}
\end{figure}

For the weak perturbation case, $Z$=-0.1 [Fig.~\ref{specZ-0.1}(a)],
the e-I interaction strength is much smaller (10 times) than the e-e
interaction. The ground-state energy (level's degeneracy g=1) is the
singlet state where the configuration with the two electrons in the
$s$-orbital is highly dominant. The first- and second-excited states
have degeneracies $g=2$ due to the symmetry of $L=\pm1$ states. Note
that at B=0 the results are independent of spins. We present the
results in the $S_z$=0 subspace. We discuss the $L=1$ case. In the
first-excited state, the most dominant configuration is (0,0;0,1),
the second configuration is (0,-1;0,2), and the last configuration
is (0,0;1,1). These three configurations and their exchange states
have coefficients with opposite signs in the wave function
($C_{(0,0;0,1)}=-C_{(0,1;0,0)}\approx0.7$,
$C_{(0,-1;0,2)}=-C_{(0,2;0,-1)}\approx0.07$, and the other
$C_{(1,1;0,0)}=-C_{(1,1;0,0)}\approx0.06$) leading to the triplet
first-excited state. The second-excited state has an opposite manner
leading to the total spin $S$=0. Plus, the second dominant
configuration and the third dominant configuration in the
second-excited state switch their relative orders (sorted in
probability) in the total wave function as compared to their orders
in the first-excited state. The highest energies in
Fig.~\ref{specZ-0.1} are the ($L$=$\pm2$,$S$=$0$) states with the
largest configuration ($0,\pm1;0,\pm1$). Other considerable
configurations are $(0,0;0,\pm2)$ and $(0,\mp1;0,\pm3)$.

In the two energy spectra presented in Fig.~\ref{specZ-0.1}, the
ground-state and the first-excited state remain as singlet and
triplet states. The fourth, fifth, and the sixth levels, which are
the ($L$=$0$,$S$=$1$) and ($L$=$\pm1$,$S$=$0$) states, exhibit a
crossing at $d=0.0008 a^{*}_{B}$. For $d<0.0008 a^{*}_{B}$, the
($L$=$0$,$S$=$1$) state has a lower energy than the 2-fold
degenerate ($L$=$\pm1$,$S$=$0$) state. The major contribution to the
wave function of the lower energy are, sorted in probability,
$(0,0;1,0)$, $(0,0;2,0)$, and $(0,0;3,0)$. Increasing
$d\ge0.0008a^{*}_{B}$ leads to the exchange between the
($L$=$0$,$S$=$1$) and ($L$=$\pm1$,$S$=$0$) states where the 2-fold
degenerate ($L$=$\pm1$,$S$=$0$) state has lower energy.

The above physics is resulted from the dominant e-I interaction when
$Z$=-1. The ($L$=0,$S$=1) and ($L$=1,$S$=0) states now have smaller
energies than the ($L$=2,$S$=0) state. Besides, it is found that the
2-fold degenerate states $(0,\pm1,0,\mp1)$, which dominate in the
third-excited energy level ($L$=$0$,$S$=$1$) for $Z=0$, have higher
energies when the impurity is present.

The positive-$Z$ case does not affect the energy spectrum as
strongly as does for the negative-$Z$ case. The reason is that the
impurity in the positive-$Z$ case induces the same type of Coulomb
interaction with the QD e-e interaction. However, such an exchange
in the relative order of the energy levels e.g. between the
(L=0,S=1) and (2,0) states (corresponding to levels $4-6$) can still
be observed [see the crossing at $d=0.44a^*_B$ in
Fig.~\ref{specPositive}]. The point $d=1.52a^*_B$ appears as a
crossing-like point but it is only an almost-zero energy gap between
the (1,0) and (0,1) states. Similar to the attractive impurity case,
a larger e-I interaction also results in a smaller energy gap
between the first-excited state and the second-excited state.

\subsection{Summary for the impurity effect on the energy spectrum of 2-electron single QDs}

The cases of negative and positive $Z$ affect the energy spectrum in
different manners. Apparently, the impurity effect for the $Z<0$
case is expected to be stronger than the $Z>0$ case. When $Z$=-1 the
e-I interaction strongly competes and dominates over the e-e
interaction. Let us take an example to illustrate the point. In the
weak perturbation regime, i.e. $|Z|=0.1$, the e-e interaction
dominates over the e-I interaction. The order of the low-level
energy spectrum, then, exhibits no difference between those two
cases $Z=\pm0.1$. Now, $|Z|$ is increased to $1$. However, the
system with negative $Z$ exhibits a stronger e-I interaction. Such a
strong effect for the attractive case $Z$=$-1$ is observed in the
exchange of the singlet and triplet states with $L$=$1$ and $L$=$0$
as seen in the inset of Fig.~\ref{specZ-0.1}(b). For the repulsive
case $Z$=$1$, this exchange is no longer observed. Only the exchange
between the triplet $L$=$0$ and singlet $L$=$2$ is found. The
repulsive e-I interaction in case $Z$=1 is manifested in the
presence of the ($L$=$2$,$S$=$0$) state which was not seen in the
lowest energy levels shown in Fig.~\ref{specZ-0.1}(b) for $Z$=-1.

\section{Coupled quantum dots}
Numerical results for the singlet-dot case were discussed in detail
mostly for $\lambda=2$ corresponding to $l_0=2a_B^*=20.6$ nm . This
value of $\lambda$ was studies based on the realistic sizes of GaAs
single QDs. However, we also theoretically discussed the results for
a typical range of $\lambda$=(1,2) as seen in
Fig.~\ref{single_Zchange_few_d}(a). For the coupled QD problem, we
used the optimized parameter set after Ref.\cite{Hu} where
$l_0=10.01$ nm ($\hbar\omega_0=11.24$ meV). Mapped on the single-dot
case, the coupled 2-dot system will have ``$\lambda$"$\approx$1.
%%%
\begin{figure}[hbtp]
\begin{center}
\vspace*{-0.5cm}\leftskip0.5cm
\includegraphics[height=8cm]{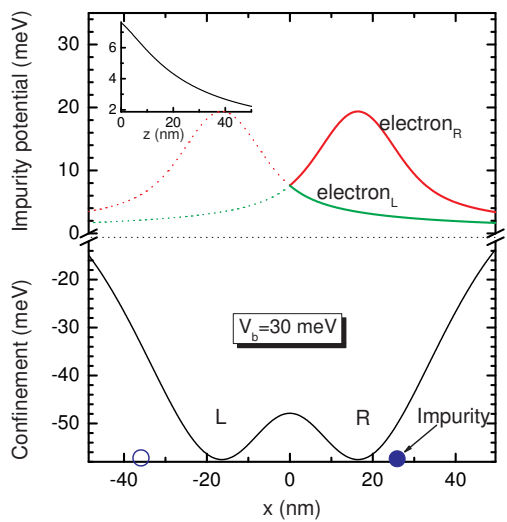}
\end{center}
\vspace{-0.8cm} \caption{Schematic plot of the confining potential
(bottom) and the charged impurity potential (top) in a coupled 2-dot
system. We assume the impurity is negatively charged with $Z_1=1$
and is located along the line connecting the two double-well minima
($\pm{a_m}$) where $\pm$, respectively, indicates the left (L)- and
the right (R)-dot. The two solid lines on top depict the impurity
potential on the two individual electrons in the two separate QDs.
The upper solid line is the potential of the electron in the right-
and the lower solid line of the electron on the left-dot,
respectively. Dotted lines on top depict the other case when the
impurity is located on the other half of the $x$-axis ($x_1<0$),
i.e. it can be found inside the left-dot. $V_b$ is taken to be 30
meV. The inset is the Coulombic potential of the impurity, located
along the growth direction, which equally repels the two separate
quantum dot electrons.} \label{schematic_Imp}
\end{figure}
The model of coupled dot system in the presence of a charged
impurity can be schematically described in Fig.~\ref{schematic_Imp}.
We plotted the impurity potential in the top panel for the case the
impurity is located along the line connecting the two minima of the
double well and in the inset for the case along the growth
direction. It is clear that the impurity effect is largest when the
impurity is at either of the two minima of the confining potential
well in the former case. Whereas the latter case has the largest
impurity effect when the impurity is found at center of the system,
i.e. $R$=0. We label the two impurity coordinates as
$\textbf{R}_{1(2)}$=(${x_{1(2)}},{y_{1(2)}},{z_{1(2)}}$).

\subsection{Singlet-triplet splitting}
\subsubsection{Impurity position dependence}
In the coupled QD system without impurity and without magnetic field
($B$=0), the two lowest energy levels are the maximally entangled
exchange spin states, respectively, the singlet $\Psi_1^S$ and the
triplet $\Psi^T$. %%
\begin{figure*}[hbtp]
\begin{center}
\vspace*{-0.5cm}\leftskip3cm
\includegraphics[width=13.2cm]{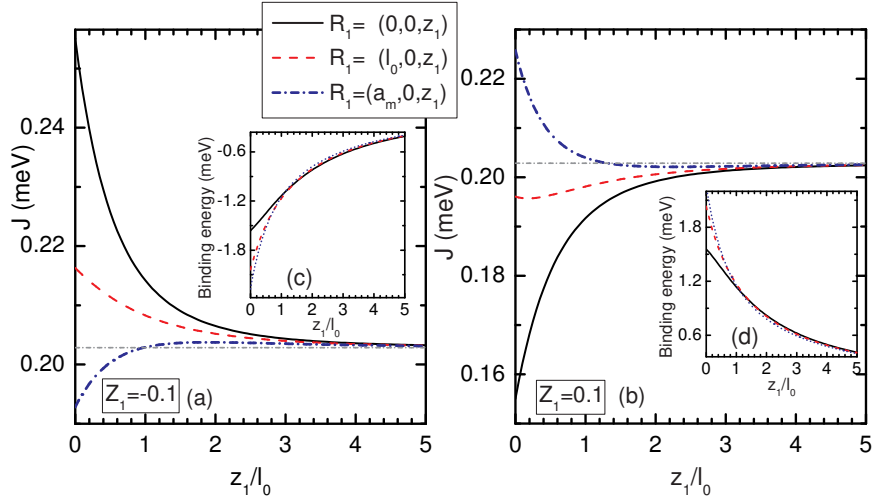}
\end{center}
\vspace{-1cm} \caption{(Color online) Singlet-triplet splitting
energy calculated as a function of $z_1$ coordinate of a single
impurity in case $Z_1$=-0.1 [(a) and (c)] and $Z_1=0.1$ [(b) and
(d)] for three different positions of the impurity $x$-coordinate
$x_1$=0 (black solid), $x_1$=$l_0$ (red dash), and $x_1$=$a_m$ -
either of the two double well minima (blue dashed-dot). $V_b$ is
taken to be 30 meV. For comparison purpose, we recall the value of
$J\approx0.204$ meV as horizontal dashed-dot line in case no
impurity is present for $B$=0. The insets (c) and (d) show the
corresponding total binding energy of the three cases plotted in
each main plot (a) and (b).} \label{double-dot_Z1M1_z1_x1=0_B=0_ST}
\end{figure*}
The next higher excited states are the linear combination of the two
doubly-occupied singlets which result in ``bonding"
($\frac{\psi_2^S+\psi_3^S}{\sqrt{2}}$) and ``anti-bonding"
($\frac{\psi_2^S-\psi_3^S}{\sqrt{2}}$) states.

Let us first consider the simplest case when the system contains
only a single charged impurity $\textbf{R}_1=(x_1,y_1,z_1)$
(effective charge $Z_1$) and the impurity plays only as a weak
perturbation to the coupled dot system. However, there will be no
restriction to the impurity location in or outside the system. Such
a system allows us to provide a direct comparison to the single-dot
case discussed earlier.

In case the e-I interaction is attractive coupling, i.e. $Z_1<0$,
the singlet-triplet exchange energy $J$ is shown in
Fig.~\ref{double-dot_Z1M1_z1_x1=0_B=0_ST}(a). When the impurity is
located along the $z$-axis [see the black solid curve in
Fig.~\ref{double-dot_Z1M1_z1_x1=0_B=0_ST}(a)], the presence of the
impurity increases the singlet-triplet spin splitting $J$ between
the two electrons as compared to the case when no impurity is
present [see horizontal grey dot curve in
Fig.~\ref{double-dot_Z1M1_z1_x1=0_B=0_ST}(a)]. This is understood as
both electrons are attracted toward the impurity. Because the
impurity equally couples to the electrons, the system favors the
antiparallel electron spin state. This type of e-I coupling reduces
the total energy of the system [negative binding energy presented as
black solid in Fig.~\ref{double-dot_Z1M1_z1_x1=0_B=0_ST}(c)]. As the
impurity is moved away from the origin [$z_1\ne0$ - still the black
solid curve in Fig.~\ref{double-dot_Z1M1_z1_x1=0_B=0_ST}(a)], $J$
will decrease. Such a decrease can be evaluated via
Eq.~(\ref{e:A_e-Iz}) (see Appendix~\ref{A:Coulomb}) as the product
of an exponential and complementary error function.

When the impurity is located along the `$z$'-direction of either the
two separate dots, e.g. of the right-dot i.e.
$\textbf{R}_1=(a_m,0,z_1)$ [see red dash curve in
Fig.~\ref{double-dot_Z1M1_z1_x1=0_B=0_ST}(a)], $J$ remains larger
than the $J$ for the case without impurity $Z_1=0$. However, $J$
behaves very differently from the above two cases $x_1=0$ and
$x_1=l_0$. The impurity not only no longer attracts equally the two
QD electrons (similar to the $x_1=l_0$ case) but also affects the
doubly-occupied states most. $J$ in this case intersects the
splitting energy of the $Z_1=0$ case at $z_1\approx{l_0}$. Close to
the QD sample, i.e. $z_1<<l_0$, $J$ is always smaller than the
splitting energy for the without impurity case, about 10$\%$. We
reserve the detailed physical discussion around ($\pm{a_m},0$) for a
later discussion when we examine the case the impurity is located
along the $x$-axis.

We expect that a repulsive impurity induces an opposite spin order
of the two electron spin orientations: the parallel spin state. This
is illustrated in Fig.~\ref{double-dot_Z1M1_z1_x1=0_B=0_ST}(b). A
weakly repulsive impurity ($Z_1=0.1$), located along the $z$-axis
(black solid curve), has the smallest exchange energy ($\approx$
0.17 meV) when it is, apparently, at the origin: $z_1=0$. The reason
is the impurity now repels both electrons. The e-I addition energy
lifts up, $\approx$ few meV, the total energy of the coupled 2-dot
system [see the positive binding energy presented in
Fig.~\ref{double-dot_Z1M1_z1_x1=0_B=0_ST}(d)]. $J$ rapidly increases
and reaches the value of the non-impurity case as the impurity is
engineered relatively far from the origin, say $3l_0$.

The ground state consists of all three Hund-Mulliken singlets, i.e.
the total wave function $\Psi_{GS}=\{\psi_1^S,\psi_2^S,\psi_3^S\}$.
However, two doubly occupied states play a small part, which are
$\approx2\%$, to the total wave function.

The binding energy of the impurity, which is defined as the energy
difference of the system with and without a charged impurity, shown
in Figs.~\ref{double-dot_Z1M1_z1_x1=0_B=0_ST}(c) and (d) was
partially discussed above. In both cases $Z_1=\mp0.1$, around
$z_1=0$, the absolute value of the binding energy is found largest
in case the impurity is placed closer to either of the two minima of
the well ($x_1=a_m$): $2.2$ meV as compared to $2$ meV and $1.53$
meV of the $x_1=l_0$ and $x_1=0$ case, respectively. Beyond a
critical $z_1$, say $z_1>2l_0$ the most dominant binding energy case
($x_1=a_m$ - black solid curve) becomes less dominant and compatible
to the other cases $x_1=0$ and $l_0$. All three curves in
Figs.~\ref{double-dot_Z1M1_z1_x1=0_B=0_ST}(c) and (d) convert to the
situation without impurity at the large $z_1$ limit. %%
\begin{figure*}[hbtp]
\begin{center}
\vspace*{-0.5cm}\leftskip3.0cm
\includegraphics[width=13cm]{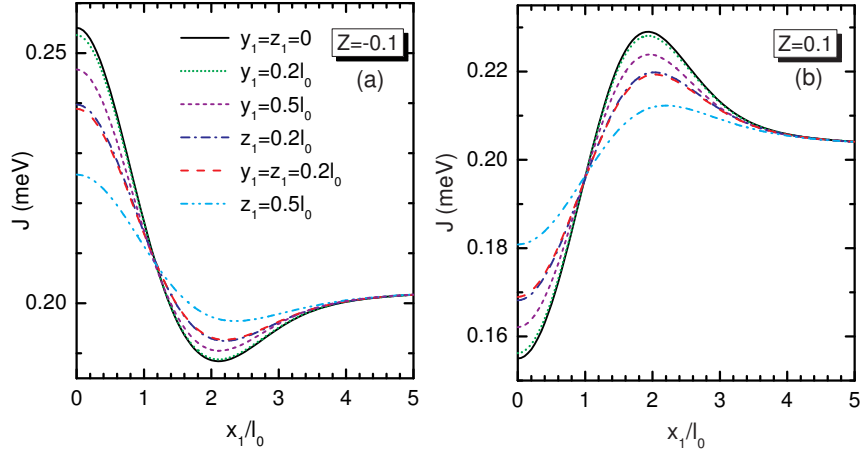}
\end{center}
\vspace{-1cm} \caption{(Color online) Singlet-triplet splitting
energy as a function of the $x$-coordinate of the impurity position
for (a) $Z_1=-0.1$ and (b) $Z_1=0.1$ for $V_b$=30 meV. The $y$- and
$z$-coordinates of the impurity position, $R_y$ and $R_z$, are
varied from the origin (black solid) to $R_y=0.2l_0$ (green dotted),
$R_y=0.5l_0$ (violet short-dashed), $R_z=0.2l_0$ (blue dash-dotted),
$R_y=0.2l_0, R_z=0.2l_0$ (red dashed), and $R_z=0.5l_0$
(dash-dot-dotted). The omitted coordinates are implied to be 0. We
cover many different possible positions of the impurity such that:
it can be along the $x$-axis or along the $y$-axis, or around the
either the two minima with both $x_1$ and $y_1$ coordinates slightly
changed while $z_1=0$, etc. Notice the case where the impurity is
shifted along the $y$-axis. This shift does not add much physics to
the system. As for example, see the insubstantial difference between
the exchange energies in the black solid ($y_1=x_1=0$) and green dot
($y_1=0.2l_0$) curves, or between the blue dashed-dot ($z_1=0.2l_0$)
and red dash ($y_1=z_1=0.2l_0$) curves, etc, presented in both (a)
and (b).} \label{double-dot_Z1M1_x1_B=0_ST}
\end{figure*}

Now, we consider the case when the impurity is found inside the QD.
In particular, we discuss the impurity effect when the impurity is
found on the $x$- or $y$-axis, along which the confining potential
is constructed.

Because the $Z_1=\mp0.1$ case was found to weakly affect the quantum
dot qubits and that the doubly-occupied states have very small
contributions to the singlet-triplet splitting, we can now use
Heitler-London model to analytically check our numerical results. We
found qualitative agreement between the results presented in e.g.
Fig.~\ref{double-dot_Z1M1_x1_B=0_ST} and the analytical results
shown in Eq.~(\ref{e:JHL}), in particular the minimum (maximum) in
$J$ for $Z$=-0.1 (0.1). Details are collected in
Appendix~\ref{A:HL}.

As signified in the top plot in Fig.~\ref{double-dot_potential}, the
absolute value of the effective e-I coupling exhibits a maximum at
the either of the two well minima and decreases rapidly as the
impurity position is out of the minima [analytical matrix elements
are presented in Eq.~(\ref{e:A_Ixy}) in Appendix~\ref{A:Coulomb}].
However, the overlap between the two coupled QDs has a maximum at
the origin (see e.g. Ref. \cite{Sausa}). In the weak impurity
perturbation, i.e. $Z_1<<1$, these two terms compete with each
other. As a result, the singlet-triplet spin splitting $J$ for
$Z_1=-0.1$ has the impurity position dependence as shown in
Fig.~\ref{double-dot_Z1M1_x1_B=0_ST}. $J$ has a maximum at the
origin where the overlap between the two coupled dots is largest and
the binding energy exhibits a maximum around $x_1=1.6l_0$ which is
identical to the quasi-bottom positions of the left and the right
dots $\pm{a}$. As compared to the case the impurity is located along
the $z$-direction, the considered case has a minimum around the
point $x_1\approx2l_0$ (see Fig.~\ref{double-dot_Z1M1_x1_B=0_ST})
which is identical to the analytical minimum obtained using
Heitler-London approximation (see Appendix~\ref{A:HL}). We note that
such a minimum is obtained for an \textit{attractive} impurity which
has a positive charge $Z_1<0$ and $|Z_1|<<1$. Moving the impurity
out of the dot system (blue dashed-dot, red dash, and cyan
dash-dotted dot in Fig.~\ref{double-dot_Z1M1_x1_B=0_ST}) will lead
to a decrease in $J$. This situation can be considered as the
coupled QDs interacting with charged impurities found close to the
surface during the growth process. In particular, the case that the
impurity $y$-coordinate is displaced to $0<y_0/l_0<<1$ has only
small difference in $J$ from the case the impurity is positioned
exactly along the $x$-axis, $y_0=0$ (see the green short-dot curves
and solid black curves in Fig.~\ref{double-dot_Z1M1_x1_B=0_ST}).
%%%
\begin{figure}[hbtp]
\begin{center}
\vspace*{-0.5cm}\leftskip-0.5cm
\includegraphics[width=10.3cm]{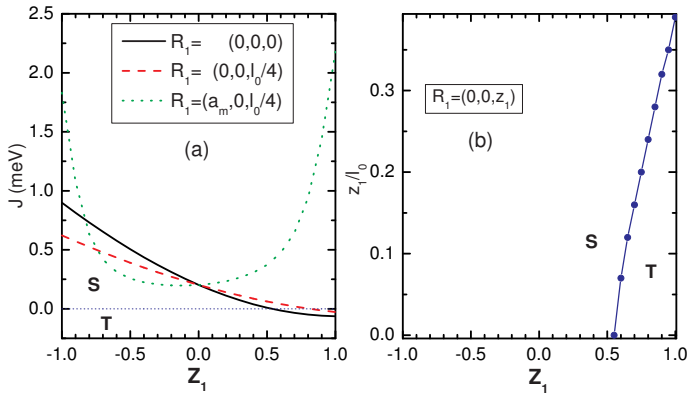}
\end{center}
\vspace{-1cm} \caption{(Color online) (a) Singlet-triplet spin
splitting $J$ as a function of the impurity effective charge $Z_1$
within the range $Z_1=(-1,1)$ for three different positions of the
impurity along the $z$-axis and (b) the singlet-triplet
state-diagram as a function of the impurity charge $Z_1$ and its
position along the $z$-axis. $V_b$=30 meV. S and T refer,
respectively, to the singlet and triplet state. When the impurity is
positioned close to the origin [$z_1$=0 (black solid) and
$z_1$=$0.25l_0$ (red dash) in (a)], the system exhibits a
triplet-singlet transition at $Z_1=0.54$ and 0.8, respectively.
Green dot curve is added for reference purpose which shows that as
the impurity is found at (or close) to either left- or right-minimum
of the well the system is always the singlet.}
\label{double-dot_Zchange+PS_z_J}
\end{figure}

\subsubsection{Impurity charge dependence}
The above critical point $x_1\approx 2l_0$ at which the
singlet-triplet splitting exhibits a maximum/minimum (see
Fig.~\ref{double-dot_Z1M1_x1_B=0_ST}) depends on the effective
impurity charge $Z_1$ and the well barrier height $V_b$. In this
subsection, we examine the $Z_1$ dependence of the singlet-triplet
spin splitting $J$.

Besides, we notice the physics sampled around the origin in
Fig.~\ref{double-dot_Z1M1_z1_x1=0_B=0_ST}(d) and
~\ref{double-dot_Z1M1_x1_B=0_ST}(b) where the presence of a
repulsive charged impurity significantly lowers the singlet-triplet
spin splitting $J$. It is expected that by further increasing the
impurity charge $Z_1$, the system can be visited in the triplet
state, i.e. $J<0$.

\begin{figure*}[hbtp]
\begin{center}
\vspace*{-0.5cm}\leftskip2cm
\includegraphics[height=7cm]{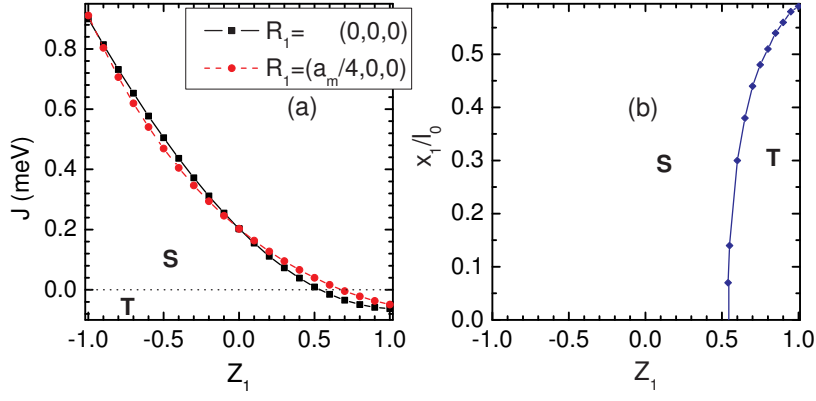}
\end{center}
\vspace{-1cm} \caption{(Color online) (a) Singlet-triplet spin
splitting of a coupled 2-dot system as a function of the effective
charge $Z_1$ within the range $-1<Z_1<1$ for $x_1=0$ (full squares
with black solid) and $x_1=a_m/4$ (full circles with red solid). (b)
Singlet-triplet state-diagram plotted in the
impurity-position$-$effective-charge $R_1-Z_1$ plane. $V_b$ is taken
to be 30 meV. Horizontal dot line in (a) is used to clarify the
triplet-singlet transition. In the region close to the center of the
system, at $\approx0.07l_0$ (about $0.04a_m$), a relatively large
repulsive impurity potential, say $Z_1=0.54$, is enough to induce a
triplet-singlet transition [see the state-diagram (b)]. The stronger
the effective charge, the broader the triplet region. Consequently,
when moving the impurity close to the right bottom it requires a
larger $Z_1$ to observe the triplet-singlet transition occur. S and
T stands, respectively, for the singlet and triplet state.}
\label{double-dot_Z1MP_Vb30_x_J1}
\end{figure*}

Such triplet-singlet transition occurs as $Z_1$ is larger than
$\approx$ 0.55 in both cases the impurity is located along the
$z$-axis and along the $x$-axis as shown, respectively, in
Figs.~\ref{double-dot_Zchange+PS_z_J} and
~\ref{double-dot_Z1MP_Vb30_x_J1}. Our calculations for the
attractive coupling case $Z_1<0$ show that there is no
triplet-singlet transition (see
Figs.~\ref{double-dot_Zchange+PS_z_J} and
~\ref{double-dot_Z1MP_Vb30_x_J1}) due to the fact that a positively
charged impurity attracts both electrons therefore the favored spin
state is always the singlet. %In a latter discussion on the strong
%impurity-effect on the energy spectrum, we will focus on other
%effect than only the triplet-singlet transition for $Z_1<0$.

The triplet-singlet transition is further explored for different
negative charges $Z_1>0$ as seen in both
Figs.~\ref{double-dot_Zchange+PS_z_J}(a) and
~\ref{double-dot_Z1MP_Vb30_x_J1}(a) for different impurity
locations: at the center [black solid in
Fig.~\ref{double-dot_Zchange+PS_z_J}(a) and full rectangles with
black solid in ~\ref{double-dot_Z1MP_Vb30_x_J1}(a)] and off-center -
slightly away from the origin. We only examine the triplet-singlet
transition for the systems with the e-I coupling smaller or
compatible to the e-e interaction. In fact, the limit of $Z_1=\pm1$
does not bear much physical meaning. However, we theoretically
examine that limit to support a complete understanding of the
effective e-I strength on the exchange electron qubits in zero
$B$-field.
%%%

It is undoubted that when the impurity equally interacts with the
two quantum-dot electrons and the e-I interaction strength is
considerable the triplet ground state occurs at a smaller $Z_1$, as
compared to the unequal electron(s)-impurity interaction case. This
argument is clarified in e.g.
Fig.~\ref{double-dot_Z1MP_Vb30_x_J1}(a) where the singlet state
occurs by increasing $Z_1$ to $\approx 0.54$ for $R_1=0$ and further
up to $0.7$ for $R_1=a_m/4$ ($\approx 0.43l_0$). On the other hand,
at a certain strong enough $Z_1$, the triplet state in case the
impurity is located along the $x$-axis remains longer than in case
the impurity is located along the $z$-axis. For example, for
$Z_1=1$, we obtain that the triplet state stays up to
$x_1\approx0.6l_0$ while it is found only up to $z_1\approx0.46l_0$.
It is worth noting that at a higher $V_b=35$ meV, due to the
influence of the impurity potential, the triplet state remains up to
a larger $x_1$ in comparison to the 30 meV case. For example, such a
triplet-singlet transition is obtained at $x_1=0.42l_0$ and
$0.72l_0$ for $Z_1$=0.6 and 1, respectively.

When the impurity is located at the right well bottom for
$Z_1=\mp1$, the ground-state is always a singlet however the
maximally entangled component $\psi_1^S$ is replaced by one of the
two double-occupied components $\{\psi_{2}^S,\psi_{3}^S\}$ as the
major contribution to the total wave function.

We summarize in Figs.~\ref{double-dot_Zchange+PS_z_J}(b) and
~\ref{double-dot_Z1MP_Vb30_x_J1}(b) the occurrence of the triplet or
singlet as the ground state when changing the impurity position
$R_1$ and its effective charge $Z_1$ for $V_b=30$ meV. The triplet
state starts to occur when $Z_1$ is increased to $\approx 0.54$ at
$R_1=0$. The largest triplet region [$R_1=(0,0.38l_0)$ in
Fig.~\ref{double-dot_Zchange+PS_z_J}(b) and ($0,0.6l_0$) in Fig.
~\ref{double-dot_Z1MP_Vb30_x_J1}(b)] is, apparently, seen for the
largest considered $Z_1$ (=1). In particular, close to the origin,
the triplet region is rapidly shortened. For example, for $Z_1$=0.6
the triplet region is in $R_1=(0,0.3)$ while such a region squeezes
to $(0,0.07)$ as $Z_1$ decreases to $0.54$. Note that we show the
singlet-triplet state-diagram on the right half of the $x$-
($z$-)axis. As the impurity is positioned along the other half of
the $x$- ($z$-)axis, the state-diagram is found similar.

\subsubsection{Inter-dot separation dependence}
\begin{figure}[hbtp]
\begin{center}
\vspace*{-0.5cm}\leftskip-0.4cm
\includegraphics[width=8.5cm]{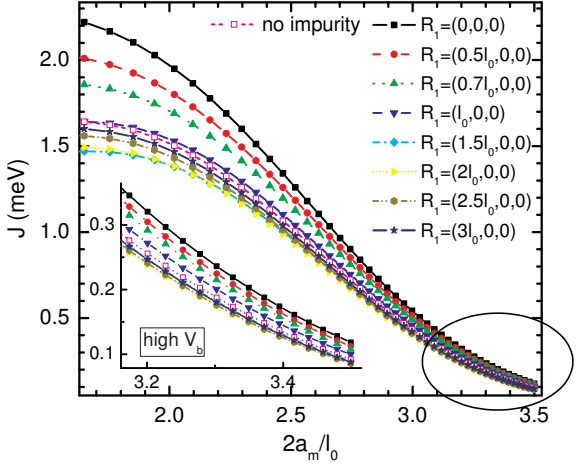}
\end{center}
\vspace{-0.9cm} \caption{(Color online) Singlet-triplet splitting
energy $J$ calculated as a function of the inter-dot separation
$2a_m$ for different positions of the impurity when it is located
along the $x$-axis for $Z_1$=-0.1, $Z_2$=0. We added data in case no
impurity is present as the open squares with magenta dash line.
Because $V_b$ changes (from 35 meV down to 13 meV) effectively
controlling the inter-dot separation, i.e. $a_m$, while the impurity
position should be fixed during the $V_b$ modification, we examine
different locations of the impurity. Several examples of
corresponding ($V_b,2a_m$) are ($35$ meV$, 3.5l_0$), ($30$ meV$,
3.3l_0$), and ($25$ meV$, 2l_0$). Inset is the magnification of the
region sampled by the big open circle for high-$V_b$-limit. }
\label{double-dot_Z1M1_center_B=0_Vbchanged_ST}
\end{figure}
%%
%%%
\begin{figure}[hbtp]
\begin{center}
\vspace*{-0.5cm}\leftskip0.1cm
\includegraphics[width=8.5cm]{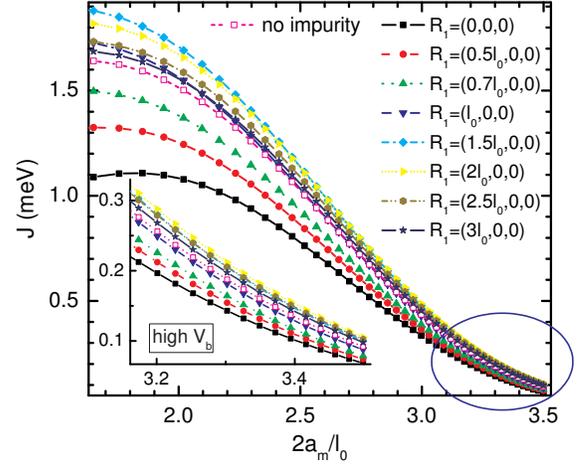}
\end{center}
\vspace{-0.9cm} \caption{(Color online) The same plot as
Fig.~\ref{double-dot_Z1M1_center_B=0_Vbchanged_ST} for $Z_1=0.1$. As
the impurity position is engineered further away from the origin,
the double-dot system reveals crossings between different $J$ for
different $R_1$.} \label{double-dot_Vbchange_Z1=0.1_J}
\end{figure}

In this subsection, we study the inter-dot dependence of the
singlet-triplet splitting $J$. First, we place the impurity at the
center of the dot system, i.e. $\textbf{R}_1=(0,0,0)$, and tune the
barrier height from $V_b=35$ meV down to $13$ meV. In the meantime,
the inter-dot separation ($2a_m$) will decrease from $2a_m=35$ nm
($\approx 3.5 l_0$) to $17$ nm ($\approx 1.6l_0$). It is worth
noting that the case that the impurity is placed exactly at either
of the two well bottoms does not have any physical meaning in this
case because the bottoms of the double-well change concomitantly
with $V_b$ changing (see Fig.~\ref{double-dot_potential}). In fact,
the impurity position should be fixed thus its position only varies
relatively with respect to the double-well minima of different-$V_b$
systems. Consequently, different-$V_b$ 2-dot systems exhibit various
impurity effect on $J$. We obtain the inter-dot separation
dependence of the singlet-triplet spin splitting $J$ for $Z_1=-0.1$
in Fig.~\ref{double-dot_Z1M1_center_B=0_Vbchanged_ST} and $Z_1$=0.1
in Fig.~\ref{double-dot_Vbchange_Z1=0.1_J}.

For $Z_1=-0.1$, the splitting energy $J$ becomes smaller as the
inter-dot separation increases (see
Fig.~\ref{double-dot_Z1M1_center_B=0_Vbchanged_ST}). However, $J$ is
found larger than that of the case without charged impurity as
illustrated in e.g. the full rectangles with black solid line
($R_1=0$) and open rectangles with magenta short dash (no impurity)
in Fig.~\ref{double-dot_Z1M1_center_B=0_Vbchanged_ST}. In this case,
this means that the potential well height is ``weakened" by the
attractive impurity which attracts the electrons toward and
therefore supports the anti-parallel spin interaction of the two
electrons. Displacing the impurity away from the center of the
double-dot system (full rectangles) along the $x$-axis results in
changes as seen in a series of $J$ presented as full circles with
red solid (for ${R}_1=0.5l_0$), full up-triangles with green dot
(for ${R}_1=0.7l_0$), full down-triangles with blue solid (for
${R}_1=l_0$), full rhombuses with  dashed-dot (for ${R}_1=1.5l_0$),
full right-triangles with yellow short-dot (for ${R}_1=2l_0$), full
circles with dash-dotted-dot (for ${R}_1=2.5l_0$), and full stars
with solid (for ${R}_1=3l_0$). First, we analyze the data for the
lowest case of $V_b=13$ meV (corresponding to $2a_m\approx1.65l_0$).
The minimum in $J$ for this case is found at
$x_1\approx1.3l_0$\cite{note1} while the maximal value in $J$ is
always found at the origin $R_1=0$. Moreover, the system tends to
covert to the no-impurity situation when the charged impurity is
engineered far enough from the center of the dot system. Therefore,
from the $R=0$-case to the $R=0.7l_0$-case we obtained such a
decrease in $J$ from $\approx 2.12$ meV to $1.87$ meV but for
$R=l_0$ we obtained $J$ almost identical to the $J$ of a similar
system but without impurity. Note that this impurity position $l_0$
is considered still close to the right-bottom of the well barrier.
This feature is similar to the physics around the impurity position
$R\approx1.2l_0$ for $V_b=30$ meV as previously examined in full
squares with black solid in Fig.~\ref{double-dot_Z1M1_x1_B=0_ST}.
Further increasing $x_1$ to $1.5l_0$ and $2l_0$ (as seem in the full
rhombuses with cyan dashed-dot and full right-triangles with
short-dot, respectively) we obtain the increase back in the latter
case. This is found due to the fact that both $1.5l_0$ and $2l_0$
are located on the right-hand-side of the minimal point in $J$. The
latter case exhibits an increase because the system to this extent
tends to convert to the non-impurity case. In the furthermost case
$R_1=3l_0$ (full stars with solid line) $J$ is found closer to that
of the non-impurity case which has higher energy than the cases
$R_1=1,5l_0, 2l_0,$ and $2.5l_0$.

Such a shift in the maximum of $J$ by changing $V_b$ can be seen in
Fig.~\ref{double-dot_Z1M1_center_B=0_Vbchanged_ST} and its inset for
the high-$V_b$ limit where the furthermost impurity positions
$R_1=1.5l_0, 2l_0, 2.5l_0,$ and $3l_0$ stay closer to the
non-impurity case in comparison to the small-$V_b$ cases. This is
related to the fact, which was mentioned above, that the relative
distance of the impurity position to the minimal point of $J$ varies
where the impurity can be found either in the left- or the
right-hand-side of the minimal $J$ as $V_b$ changes. For $V_b=30$
meV (corresponding to $2a_m\approx3.3l_0$), from the center-$R_1=0$-
to the $l_0$-case the singlet-triplet spin splitting stays higher in
energy while the others ($R_1=1.5l_0, 2l_0, 2.5l_0$, and $3l_0$)
have lower $J$ than the non-impurity case. The minimal point for
this system is $\approx2l_0$. Therefore, the $R_1=1.5l_0$-case has a
higher $J$ than the $R=2l_0$-case which was opposite in the previous
case $V_b=13$ meV. As a consequence, there appear crossings in $J$
for different $R_1$. Obvious crossings are seen for the cases
$R_1=1.5l_0, 2l_0, 2.5l_0$, and $3l_0$. For example, $J$ for the
$R_1=0.5l_0$-case (full rhombuses with cyan dashed-dot) stays lower
in energy however it changes as $V_b$ varies and is found higher (or
equal) in energy than the $J$ in the others (see the inset as a
magnification for the high-$V_b$-limit).

Apparently, when placing the impurity close to the origin we always
obtain a larger singlet-triplet spin splitting as compared to the
non-impurity case. The largest case has $\Delta{J}\approx 0.6$ meV
for $R_1=0$ and $V_b=13$ meV. It is worth noting that a larger $V_b$
results in a smaller difference in $J$ which can be understood as a
lower ``tunneling" rate of the two electrons in the two separate
QDs. Quantitatively, such a decrease can be roughly estimated by
Eq.~(\ref{e:JHL})

It was made clear in the presence of an attractive charged impurity
the e-I coupling increases the ``tunneling" rate between the two
dots. Positioning the impurity at different locations inside the
coupled 2-dot system, in particular around the $(x_1\approx2l_0,0)$
point, gives rise to different $V_b$-dependent singlet-triplet
splitting which has crossings between different $R_1$-curves. %%
\begin{figure}[hbtp]
\begin{center}
\vspace*{-0.5cm}\leftskip0.2cm
\includegraphics[width=7.1cm]{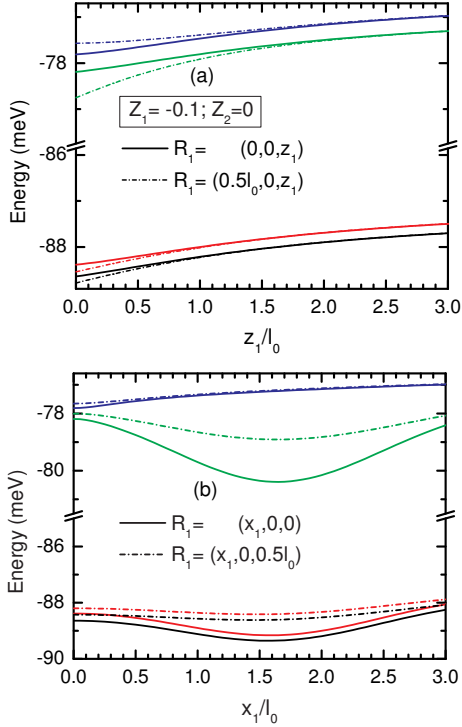}
\end{center}
\vspace{-0.8cm} \caption{(Color online) Energy spectrum of the QD
system studied (a) in Fig.~\ref{double-dot_Z1M1_z1_x1=0_B=0_ST}(a)
and (b) in Fig.~\ref{double-dot_Z1M1_x1_B=0_ST}(a), respectively, as
a function of the impurity position (along the $z$ and the $x$-axis)
for $Z_1=-0.1, Z_2=0$. $V_b=30$ meV. Because there is no
crossing/anti-crossing in the energy spectrum, we use the same style
of line for a certain $R_1$-dependence. In (a), solid curves depict
the impurity-position dependence when the impurity is located along
the $z$-axis and dash-dotted curves depict the situation when the
impurity is outside the double-dot system [along the  line ($y_1=0;
x_1=0.5l_0$)]. In (b), solid curves depict the case $R_1$ is changed
along the $x$-axis and dash-dotted curves depict the situation where
the impurity is outside the double-dot system but along the
$x$-direction. The latter case  of (a) with $x_1=0.5l_0$ (dashed-dot
curves) reveals a stronger impurity effect on the coupled qubits in
comparison to the case when the impurity is on the $z$-axis [solid
curves in (a)]. The $R_1=(0,0,z_1)$-case induces equal exchange
coupling to the two electrons mostly found on the individual dots.
The latter case of (b), in opposite, has much weakened impurity
effect on the double-dot system in comparison to the case where the
impurity is found on the line connecting the two confining potential
minima. } \label{double-dot_Z1M1_twox1_B=0_trum}
\end{figure}
As illustrated now in Fig.~\ref{double-dot_Vbchange_Z1=0.1_J} for
$Z_1=0.1$ case, $J$ in the $R_1=0$ case (full squares with black
solid) stays lowest in energy and the $J$-difference as compared to
the non-impurity case is found largest. For example, $\Delta
J\approx 5.5$ meV for $V_b=13$ meV. $J$ in the furthermost case
$R_1=3l_0$ (full stars with navy solid) stays slightly above the
non-impurity curve (open squares with magenta short-dot). This
implies that such an impurity position is on the right-hand-side of
the maximum point in $J$ for all the considered $V_b$ [$\in$(13, 35)
meV] in Fig.~\ref{double-dot_Vbchange_Z1=0.1_J}. It is found
opposite to the $Z_1=-0.1$ case where the $R_1=3l_0$-$J$-curve stays
slightly lower than the non-impurity one. Crossings between
different $R_1$-dependent $J$-curves are still obtained in the
current case $Z_1=0.1$, e.g. the one around $2a_m=2.1l_0$
(corresponding to $V_b=15.5$ meV) between the $l_0$- (full
down-triangles with blue solid) and the $3l_0$-$J$-curves, or the
one at $2a_m\approx2.61l_0$ ($V_b=20$ meV) between the $1.5l_0$-
(full rhombuses with cyan dashed-dot) and the $2l_0$-$J$-curve (see
also the inset of Fig.~\ref{double-dot_Vbchange_Z1=0.1_J}), etc. %%

It is important to note that the $|Z_1|=0.1$ case studied so far has
not exhibited any crossing between the singlet and triplet states,
i.e. the triplet state always stays higher in energy than the
singlet state. The presence of an attractive impurity increases the
singlet-triplet splitting opposed to the effect seen for a repulsive
impurity. The extreme behavior of $J$ is obtained at either of the
two minima of the potential well. An attractive impurity seems to
increase the ``tunnel" rate of the coupled double-dot system.

\subsection{Energy spectrum}
In the single QDs as discussed in Sec. III, impurity effect on the
energy spectrum of the system is substantial as e-I interaction is
competitive to the e-e interaction. We obtained different crossings
between low-lying excited states and anti-crossings with varying
energy gaps. If such a similar effect is found in the coupled double
QD system, the question ``whether quantum operations in the coupled
dot system are affected" attracts our attention.

In order to examine the validity of using coupled dots containing
charged impurities as the basis of quantum logic gates in quantum
computation, we mainly look into the low-level energy spectrum.
\begin{figure}[hbtp]
\begin{center}
\vspace*{-0.5cm}\leftskip0.2cm
\includegraphics[width=7.0cm]{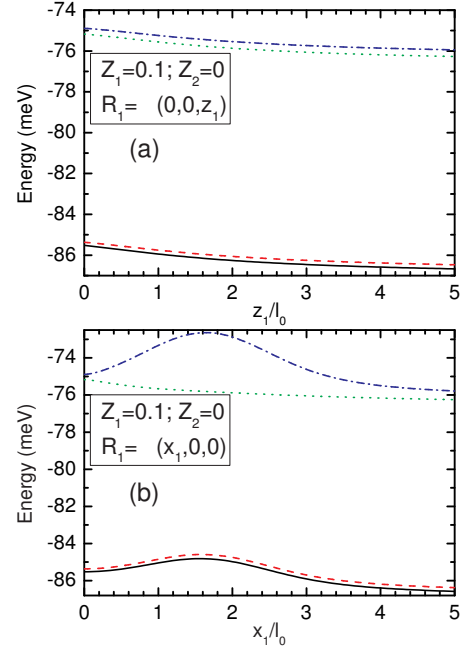}
\end{center}
\vspace{-0.8cm} \caption{(Color online) The same plot as shown in
Fig.~\ref{double-dot_Z1M1_twox1_B=0_trum} for a repulsive impurity
$Z_1=0.1$. There is a change in the relative order of the two
excited states in (b) as compared to the above case $Z_1=-0.1$ (see
Fig.~\ref{double-dot_Z1M1_twox1_B=0_trum}).}
\label{double-dot_Z1P01_Vb30_xz_trum}
\end{figure}

The energy spectrum of double QDs containing a single impurity
located at $\textbf{R}_1=(0,0,z_1)$ and $(0.5l_0,0,z_1)$,
respectively, is shown as solid and dash-dotted curve in
Fig.~\ref{double-dot_Z1M1_twox1_B=0_trum}(a). The latter case
(dash-dotted) has a larger impurity effect on the energy gap between
the first two lowest levels and the two highest energy levels which
can be understood in an earlier discussion. The energy levels bend
down to the $z_1=0$ as its $x$-coordinate $x_1\ne$ as compared to
the case $x_1=0$ because the impurity has a stronger interaction to
the electrons when it is closer to the well bottoms.

The $x_1$-dependence of the energy spectrum is studied in
Fig.~\ref{double-dot_Z1M1_twox1_B=0_trum}(b) for two different
$z_1=0$ and $z_1=l_0/2$. The minimum, discussed earlier in
Fig.~\ref{double-dot_Z1M1_x1_B=0_ST} related to the physics of
having an impurity around the well bottoms, is seen in the energy
levels of Figs.~\ref{double-dot_Z1M1_twox1_B=0_trum}(b) for $z_1=0$.
The spin splitting between the two fully-filled states is most
affected also when the impurity is at the well minima. The
probability of finding the two electrons in the right dot
(containing the impurity) increases. The absolute value of the e-I
coupling for the right (in the present case) doubly-occupied state
is maximal at $(a_m,0)$.

For $Z_1=0.1$ [see Fig.~\ref{double-dot_Z1P01_Vb30_xz_trum}(b)], it
is found opposite to the attractive $Z_1=-0.1$ case. The impurity
(in the right-dot) now tends to repel both electrons which leads to
a higher probability of finding the two electrons in the left-dot.
Consequently, the doubly-occupied state of the right-dot gains
energy via the e-I interaction and stays higher in energy (blue
dashed-dot) than the singlet state (green dot) of the left-dot. Such
a maximum around the bottom $(a_m,0)$ in the blue dashed-dot curve
in Fig.~\ref{double-dot_Z1P01_Vb30_xz_trum}(b) has similar physics
discussed earlier for $Z_1=-0.1$. Remember that in the attractive
case, the singlet-triplet spin splitting has a minimum around $2l_0$
(see Fig.~\ref{double-dot_Z1M1_x1_B=0_ST}) which is resulted due to
i) the competition between the wave function overlap of the coupled
2-dots and the impurity potential on the right-dot electron and ii)
the doubly-occupied states have relatively small contribution to the
total singlet state (the ground state). Now, the effective charge
changes its sign ($0.1$) and we obtain a maximum [see the inset of
Fig.~\ref{double-dot_Z1P01_Vb30_xz_trum}(b)]. The other left-dot
fully occupied state [green dot in
Fig.~\ref{double-dot_Z1P01_Vb30_xz_trum}(b)] only shows a decrease
in $x_1$.

Even though the presence of the impurity changes the singlet-triplet
splitting as discussed above, the energy gap $\Delta{\epsilon}$
between the highest spin state that stores information and the
lowest unwanted state, in this case the first and the second excited
states, remains much larger than the singlet-triplet splitting
energy $J$. In case $V_b=30$ meV, $J/{\Delta{\epsilon}}$ is
typically $\le 0.07$. This means that adiabatic condition is
satisfied so that higher excited states are not involved when the
system is evolved to its desired state.

\subsection{Strong electron-impurity coupling destroys
coupled qubits}

\subsubsection{Strong perturbation}
In the following, we discuss in detail the influence of a strongly
perturbative impurity on the energy spectrum for $Z_1=\mp1$. Results
for other intermediate $Z_1$ (e.g. $\mp$0.6, $\mp$0.8) are collected
in Appendix~\ref{A:Intermediate}.

We plot in Fig.~\ref{double-dot_Vb30_Z1=MP1_z_J} the
impurity-position dependence of the singlet-triplet spin splitting
$J$ when the impurity is located along the $z$-axis for (a) $Z_1=-1$
and (b) $Z_1=1$.  The former case has a similarly qualitative curve
to the case $Z_1=0.1$ with a much larger exchange energy $J$. The
latter case has a triplet-singlet transition which occurs at
$z_1=0.38l_0$ as obtained already in the singlet-triplet
state-diagram Fig.~\ref{double-dot_Zchange+PS_z_J}(b). This property
is resulted from the strong e-I interaction as $Z_1$ is increased to
1. %%
\begin{figure}[hbtp]
\begin{center}
\vspace*{-0.5cm}\leftskip-0.5cm
\includegraphics[width=10cm]{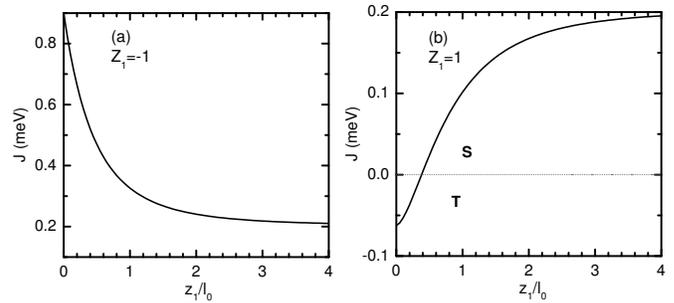}
\end{center}
\vspace{-0.8cm} \caption{(Color online) Singlet-triplet spin
splitting $J$ calculated as a function of the impurity position
along the $z$-axis for $Z_1$=-1 (a) and 1 (b). $V_b=30$ meV. S and T
stand for, respectively, singlet and triplet. The horizontal dotted
line in (b) is used to clarify the triplet and singlet states. Such
a triplet-singlet transition is obtained at $z_1=0.3l_0$ for
$Z_1$=1. } \label{double-dot_Vb30_Z1=MP1_z_J}
\end{figure}
%%
%%%%%
\begin{figure}[hbtp]
\begin{center}
\vspace*{-0.5cm}\leftskip0.2cm
\includegraphics[width=7.5cm]{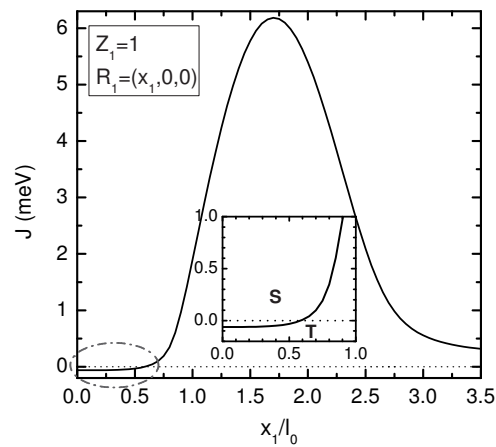}
\end{center}
\vspace{-0.8cm} \caption{(Color online) Singlet-triplet spin
splitting $J$ of a singly-doped charged impurity coupled double-dot
system as a function of the impurity position when it is located
along the $x$-axis for $Z_1=1$. $V_b=30$ meV. Inset is the
magnification of the circled region in the main plot which
highlights the triplet-singlet transition in the small $x_1$
region.} \label{double-dot_Z1MP_Vb30_x_J}
\end{figure}
%%%herehere
\begin{figure*}[hbtp]
\begin{center}
\vspace*{-0.5cm}\leftskip1.5cm
\includegraphics[width=13.7cm]{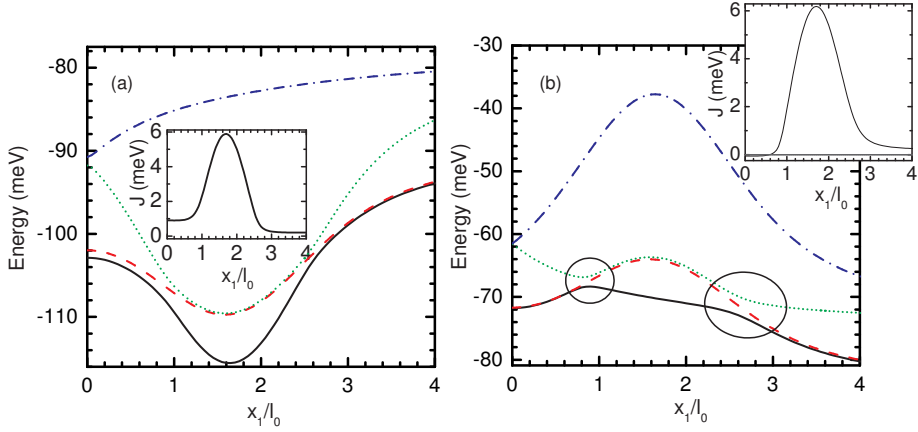}
\end{center}
\vspace{-1cm} \caption{(Color online) Energy spectrum of a coupled
QD system exhibits strong impurity effect examined as a function of
the (single) impurity position located along the $x$-axis for a (a)
positively ($Z_1$=-1) and (b) negatively ($Z_1$=1) charged impurity.
The inset in each layer is the corresponding singlet-triplet spin
splitting $J$ where both exhibit a maximum around the right-bottom
of the well. Different from the attractive case $Z_1=-1$ in (a), the
repulsive case $Z_1=1$ in (b) overcomes a triplet-singlet transition
at $\approx0.6l_0$ [see the inset of (b)]. The green dot curve in
(a) represents either the $\psi_3^S$ singlet or the $\psi_1^S$
singlet (see discussion in text) while the green dot in (b)
represents either the $\psi_2^S$ singlet or the $\psi_1^S$ singlet.
Red dash in both (a) and (b) is the triplet state. Circles that
sample the anti-crossings indicate the transition of $\psi_1^S$ as
the ground-state to the first excited state and the transition back
to the ground state as $x_1$ increases. Anti-crossings and almost
zero energy gap between the first and second excited states in both
(a) and (b) are the evidences to claim that the coupled qubits are
``destroyed" and the adiabatical condition is violated. }
\label{double-dot_Z1MP1_Vb30_x_trum}
\end{figure*}

For the case the impurity is located along the $x$-axis and $Z_1=1$,
we obtain in Fig.~\ref{double-dot_Z1MP_Vb30_x_J} and its inset the
entire triplet state for the case the impurity is engineered at the
origin or very close to the origin, say $R_1<0.6l_0$ (see the
inset). The $R_1=0$-case has the impurity that equally repels the
two electrons in the two individual QDs and the impurity is kept
distant from the two well minima. As a result, the favored state
becomes the triplet with two electron spins aligned parallel to each
other. Moving the impurity off-center means that there appears a
bias in the impurity-electron coupling with the two electrons. Such
a triplet state becomes weakened and disappeared when the bias
increases to its maximum for the case the impurity is found around
the bottoms of the confining potential.

\subsubsection{Destroy of coupled qubits}
%%%%
\begin{figure}[hbtp]
\begin{center}
\vspace*{-0.5cm}\leftskip0.1cm
\includegraphics[width=8.39cm]{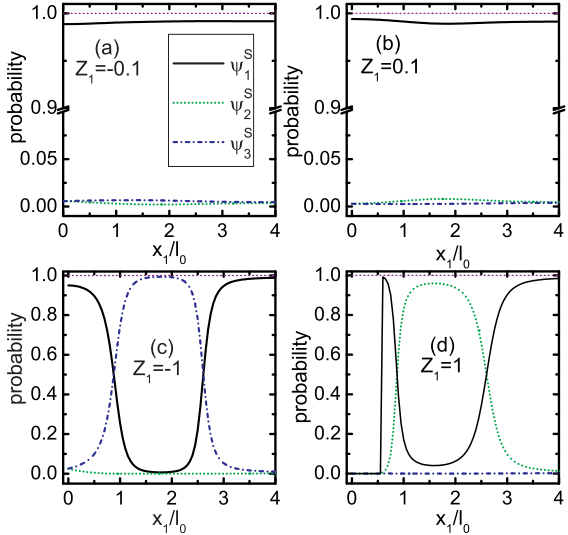}
\end{center}
\vspace{-0.8cm} \caption{(Color online) Probability of finding the
three singlet states as defined in Eq. (\ref{e:Bell}) in the ground
state as a function of the impurity position for different $Z_1$ for
weak [$Z_1$=-0.1(a) and 0.1(b)] and strong [$Z_1$=-1(c) and 1(d)]
impurity potential. The triplet state can be the ground state and
only ``couples" (e.g. via an anti-crossing as discussed earlier in
Fig.~\ref{double-dot_Z1MP1_Vb30_x_trum}) but does not mix with the
singlet states therefore the probability of finding such state is
always 1. The impurity position is $\bf{R}$$_1$=$(x_1,0,0)$. Top
dash-dotted horizontal line corresponds to the probability of
finding the $\Psi_1^S$ state in case without impurity for B=0. }
\label{double-dot_Vb30_Z1_prob}
\end{figure}
%%%

Impurity effect of small effective charges shown in
Figs.~\ref{double-dot_Z1M1_x1_B=0_ST} and
~\ref{double-dot_Z1P01_Vb30_xz_trum} as discussed above for
$Z_1=\mp0.1$ results in a relatively small coupling between the
three Hund-Mulliken singlets in ground-state singlet.
Quantitatively, the major ($\approx98\%$) component is the
$\psi^S_1$ therefore the mixing of $\psi^S_1$ with the other two
singlets $\psi^S_{2,3}$, with the probability about $1\%$ each, can
be neglected. The considerable impurity effect is observed in the
excited singlets where such full bonding- and anti-bonding states
are no longer found at B=0 when the impurity is found in either of
the two individual QDs. Plus, the probability of finding the
electrons in a fully-occupied state is increased.

In the next examination, we study the case when the coupling between
the electron and the impurity is compatible to the electron pair
Coulomb interaction by increasing the absolute effective impurity
charge $Z_1$ to $\pm1$. The energy spectrum are now shown in
Figs.~\ref{double-dot_Z1MP1_Vb30_x_trum}(a) and (b). For the
attractive impurity case [see
Fig.~\ref{double-dot_Z1MP1_Vb30_x_trum}(a)], the triplet remains as
the first excited state while the ground state now mixes three
singlets with the leading term varies either the spin-exchange
singlet, $\Psi_1^S$, or the right-dot doubly-occupied state
$\psi^S_3$. The other singlet $\psi^S_2$ (left-dot) has slightly
larger than zero contribution.

Particularly, the singlet-triplet splitting in case $Z_1=-1$ changes
mostly differently in comparison to the weak impurity potential case
$Z_1=-0.1$. $J$ now has somewhat similar physics to the case $Z=1$
around the right bottom of the potential well. A maximum in $J$
($\approx 6$ meV) was seen on both cases as shown in the insets of
Figs.~\ref{double-dot_Z1MP1_Vb30_x_trum}(a) and (b).

Now, the strong e-I interaction leads to the triplet-singlet
transition at $R_1\approx0.6l_0$ (as discussed already) for $Z_1=1$
and ``virtual" coupling between the three states $\psi^T$,
$\psi_1^S$, and $\psi_2^S$ with each other.

As compared to the weak perturbative case where the minor
contributions of the $\psi_{2,3}^S$ states, up to $1\%$ each, are
negligible, such contributions now start increasing and can exceed
$50\%$ and even larger for the doubly occupied singlet of the dot
containing the attractive impurity (the right-dot i.e. the
$\psi_3^S$) and of the dot without impurity (the left-dot i.e. the
$\psi_2^S$) for the repulsive case. Consequently, the
doubly-occupied state can become the ground state and the exchange
spin singlet becomes the first excited singlet. From
Fig.~\ref{double-dot_Z1MP1_Vb30_x_trum}(a), the green short dot
corresponds to the $\psi_3^S$ (major) and the blue dashed-dot the
$\psi_2^S$ (major) state in the region $R_1<0.8l_0$ and
$R_1>2.7l_0$. The region in between $0.8l_0<R_1<2.7l_0$ has the
ground state as the doubly occupied singlet $\psi_3^S$ in
Fig.~\ref{double-dot_Z1MP1_Vb30_x_trum}(a) and $\psi_2^S$ in
Fig.~\ref{double-dot_Z1MP1_Vb30_x_trum}(b). The minor components in
the first excited state in both cases, respectively, $\psi_1^S$ plus
$\psi_3^S$ and $\psi^S_1$ plus $\psi_2^S$, fluctuates around $7\%$
and $3\%$ as the impurity is located close to the center of the
system. In Fig.~\ref{double-dot_Z1MP1_Vb30_x_trum}(b) we see a
switch between the left- and right-doubly-occupied singlets where
the singlet of the right-dot (with impurity - blue dashed-dot) stays
higher in energy than the one of the left-dot (green dot).

The ``virtual" coupling between the three singlets and the triplet
addressed above is now examined in both cases $Z_1=\mp1$. At the
well right-bottom $R_1=(a_m,0)$, the triplet (red dash) ``couples"
with the first excited singlet as green dot curve [$\psi^S_3$ in
Fig.~\ref{double-dot_Z1MP1_Vb30_x_trum}(a)] and [$\psi^S_2$ in
Fig.~\ref{double-dot_Z1MP1_Vb30_x_trum}(b)]. The energy gap between
the triplet and the first excited state is found close to zero. This
virtual coupling can be understood by the two anti-crossings between
the $\psi_1^S$ and $\psi_3^S$ in
Fig.~\ref{double-dot_Z1MP1_Vb30_x_trum}(a) and between $\psi_1^S$
and $\psi_2^S$ in Fig.~\ref{double-dot_Z1MP1_Vb30_x_trum}(b).

To better understand the physics manifested in the mixing of
different singlet states as well as the ``coupling" between the
triplet and the first excited singlet as shown in
Figs.~\ref{double-dot_Z1MP1_Vb30_x_trum}(a) and (b) and the
triplet-singlet transition in case $Z_1=1$ we calculate the
probability of finding the three singlet states as a function of the
impurity position inside the double-dot system for weak and strong
e-I coupling, respectively, in
Figs.~\ref{double-dot_Vb30_Z1_prob}(a) and (b) and
Figs.~\ref{double-dot_Vb30_Z1_prob}(c) and (d) for the ground state.
We first explain for the triplet-singlet transition in case $Z_1=1$
[see Fig.~\ref{double-dot_Z1MP_Vb30_x_J}(a)] and the ``coupling" of
the first excited singlet and the triplet in case $Z_1=\mp1$ [see
Figs.~\ref{double-dot_Z1MP1_Vb30_x_trum}(a) and (b)]. The
triplet-singlet transition can be understood from
Fig.~\ref{double-dot_Vb30_Z1_prob}(d) where the region
$x_1/l_0\leq0.6$ has zero probability of finding the $\psi_1^S$
state in the ground state. It is because the triplet state becomes
the ground state and the singlet $\psi_1^S$ is the major component
of the first excited state wave function in this considered region.
The dominant of the doubly occupied singlet $\psi^S_3$ in the ground
state for $Z_1=-1$ and of the $\psi_2^S$ for $Z_1=1$ is now
specified as the region $0.8l_0<x_1<2.7l_0$ in
Figs.~\ref{double-dot_Vb30_Z1_prob}(c) and (d). This region is
identical to the maximal behavior of $J$ as seen in
Figs.~\ref{double-dot_Z1MP1_Vb30_x_trum}(a) and (b) and their two
insets. For weak impurity potential cases $Z_1=\mp0.1$ there are no
anti-crossings in the energy spectrum and the doubly-occupied
singlets remain always as the excited singlets with small
probability [see Figs.~\ref{double-dot_Vb30_Z1_prob}(a) and (b)].

\subsection{Coupled qubits perturbated by two impurities}

\begin{figure}[hbtp]
\begin{center}
\vspace*{-0.5cm}\leftskip0.6cm
\includegraphics[width=6.cm]{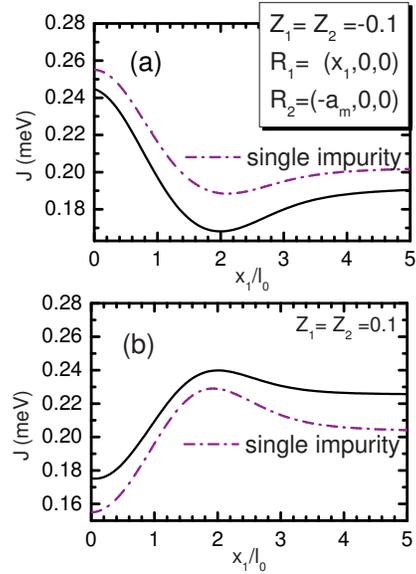}
\end{center}
\vspace{-0.8cm} \caption{(Color online) Singlet-triplet splitting
(black solid) of a coupled two-dot system with the presence of two
identical charged impurities but located asymmetry along the
$x$-axis: one is fixed at the left bottom and the other is located
along the $x>0$-axis for weak e-I coupling (a) $Z_1=Z_2=-0.1$ and
(b) $Z_1=Z_2=0.1$. $V_b$ is taken to be 30 meV. Violet dashed-dot
curves are the $J$ for the single-impurity system extracted from
Fig.~\ref{double-dot_Z1M1_x1_B=0_ST}.}
\label{double-dot_Vb30_Z1=Z2=-0.1_B=0_trum}
\end{figure}
For simplicity, we first assume that two impurities have identical
charges $Z_1=Z_2$ and one impurity is kept at one of the two minima
$\textbf{R}_2=(-a_m,0,0)$. The other impurity can be arbitrarily
located along the other side of the $x$-axis,
$\textbf{R}_1=(x_1>0,0,0)$. Under these circumstances, the two
impurities generally induce different e-I couplings and hence affect
the qubits asymmetrically except when $x_1=-x_2=a_m$. In this case,
both doubly-occupied singlets are found compatible which have
similar contribution to the two excited singlet states. Therefore,
we no longer see a minimum or maximum, depending on the sign of
$Z_1$, in the second or third excited state as obtained in case
$Z_2=0$. Such a pronounced minimum is now ``eliminated" due to the
presence of the second impurity, located at the other bottom (left)
of the 2-dot system, which induces a competing e-I coupling. The
singlet-triplet spin splitting $J$ has maximal behavior when the
impurity is found close and at the bottom of the right-dot. %%
\begin{figure}[hbtp]
\begin{center}
\vspace*{-0.5cm}\leftskip-0.0cm
\includegraphics[width=7.2cm]{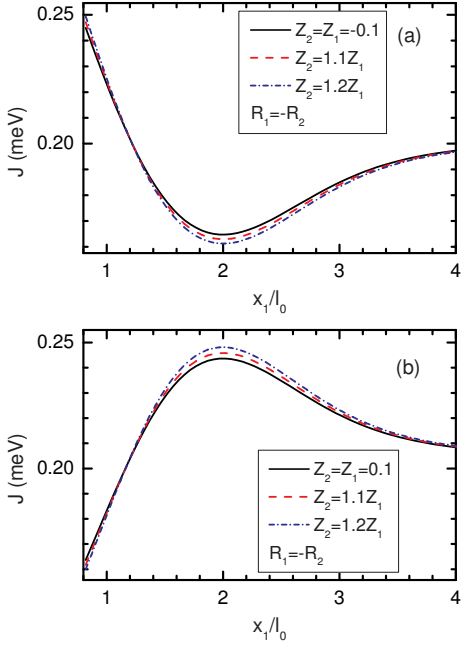}
\end{center}
\vspace{-0.8cm} \caption{(Color online) Singlet-triplet spin
splitting $J$ as a function of the impurity position in case the two
impurities are symmetrically located, $\bf{R}$$_1$=-$\bf{R}$$_2$,
but they are charged slightly differently (a) both impurities are
positively charged and (b) both impurities are negatively charged
with $Z_1=\pm0.1$ and $Z_2=Z_1$ (black solid), $1.1Z_1$ (red dash),
and $1.2Z_1$ (blue dash-dotted). $V_b$ is taken to be 30 meV. $V_b$
is fixed at 30 meV which corresponds to $a_m\approx1.64l_0$. Note
that the system converts to the situation without impurity at a much
larger $x_1$ than the singly-doped impurity case (previously seen
around $3.5l_0$). In this plot, at $x_1=4l_0$ the system does not
convert to the case without impurity (e.g. $\Delta{J}\approx$0.01
meV) and the total binding energy $\Delta{E}\approx$0.8 meV for
$Z_1=-0.1$ [black solid in (a)]. }
\label{double-dot_Vb30_changeR1=-R2_B=0_J}
\end{figure}

Another way to study the asymmetric impurity effect in the
considered double-dot system with two impurities is to adjust the
effective impurity charges with respect to each other while their
positions are kept symmetrical to the origin, i.e. $Z_1/Z_2\neq1$
while $\textbf{R}_1$=-$\textbf{R}_2$. We compute the singlet-triplet
spin splitting for three different values of $Z_2/Z_1=1, 1.1,$ and
1.2 while $R_{1,2}$ changes for both positively and negatively
charged cases in Fig.~\ref{double-dot_Vb30_changeR1=-R2_B=0_J}(a)
and (b), respectively. The results shown in
Fig.~\ref{double-dot_Vb30_changeR1=-R2_B=0_J} are obtained for
$Z_2/Z_1$ slightly different than unit to examine the asymmetry
arisen due to the dissimilarity in impurity charges as compared to
the asymmetry due to the impurity positions discussed earlier in
Fig.~\ref{double-dot_Vb30_Z1=Z2=-0.1_B=0_trum} for $Z_1=Z_2$.

Apparently, if one replaces one of the two identical impurities by
an opposite charge to the other one can revisit the ``asymmetric"
phenomenon between the two doubly occupied singlets in the two dots
which now can be enhanced as a double. Our further calculations show
that if $Z_1$=-$Z_2$=-$0.1$ and
$\textbf{R}_1=-\textbf{R}_2=(-a_m,0,0)$, for $V_b$=30 meV the two
doubly-occupied states have a significantly large energy gap and the
higher level, which is the right occupied state in this case,
exhibits a maximum around $(a_m,0)$ - one of the two potential well
minima. Generally, this considered situation can be extended to the
case that the charged impurities are different types and obey a
specified charge distribution. By examining the energy spectrum of
the coupled dot, as illustrated in the above discussion, one can
detect the impurity effect on the singlets of the double-dot system.
This can be made possible due to the distinguishability between the
total positive and the total negative charge cases. It is also
possible that both impurities are found either in the left- or
right-dot, however, not much different physics from the case with
only a single impurity will be added.

\begin{figure}[hbtp]
\begin{center}
\vspace*{-0.8cm}\leftskip-0.0cm
\includegraphics[width=7.5cm]{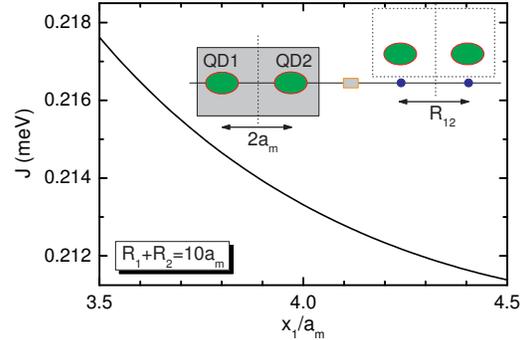}
\end{center}
\vspace{-0.5cm} \caption{(Color online) Singlet-triplet spin
splitting $J$ as a function of the impurity position in case the two
impurities (blue circles) are located relatively far from the two
potential minima, $\bf{R}$$_1$+$\bf{R}$$_2$/2=5$a_m$ and the closest
impurity to the active coupled 2-dot system (the grey box) starts at
$R_1=3.5a_m$. By fixing the center-of-mass coordinate of the two
impurities at $R_c=5a_m$, we imply such two impurities can be
considered as another coupled 2-dot system which has no tunneling
coupling with the considered 2-dot system in the grey box. That the
relative positions of each impurity $\bf{R}$$_{1,2}$ vary with
respect to the origin (0,0) of the active system can be mapped to
the problem of other 2-dot system (in open dotted box) with varying
potential barrier height. $V_b$ of the active system is taken to be
30 meV. $J$ is slightly different from that of the system without
impurity.} \label{double-dot_Vb30_special_J}
\end{figure}

The last part of this section is reserved to examine one special
case described in what follows. The coupled 2-dot system interacts
with two negatively charged impurities $Z_1=Z_2=1$, equal to the
electron charge. Now, the two impurities are engineered such that
they are far enough from the target (A) coupled 2-dot sample and
only induces electrostatic interaction to the two electrons of the
target system. We examine the case that the two impurities have
varied positions but their center-of-mass coordinates are kept
unchanged, in this case
$\textbf{R}_{12}^c=(\textbf{R}_1+\textbf{R}_2)/2=(5a_m,0,0)$. This
center-of-mass location can be imagined as another ``center" of
another coupled 2-dot system (B) where the two impurities are mostly
found at the well bottoms of the system B. That the relative
positions of individual impurities vary with respect to their
center-of-mass position can be imagined as the well barrier height
is tuned resulting in varying inter-dot separation in system B. A
and B only interact via the e-I Coulomb interaction. Now, we study
the influence of the presence of system B on the qubits of the
system A. We plot in Fig.~\ref{double-dot_Vb30_special_J} the
singlet-triplet spin splitting as a function of one impurity
position for system A. We find that $J$ decreases as the impurity
which stays closer to the system A is positioned further from the
system A. This effect is expected to be significantly enhanced if
one applies an external field to engineer the impurities in and out
the active region of the target system A. As a result, one can see
such anti-crossings as seen earlier for $Z=\mp1$ where the triplet
state stays in-between the two lowest singlets.
%%%

\section{Discussion and conclusion}

The coupling of two qubits is sensitively affected by the presence
of a charged impurity. A charged impurity which is found inside the
right- or the left-dot ``breaks down" the equality between the right
and the left doubly occupied singlet states. In case the impurity
weakly couples with the electrons, the maximally doubly-occupied
singlet of the dot containing the impurity has a smaller energy due
to the e-I coupling. This can be considered as a tool to distinguish
the doped and un-doped QD components in the coupled QD system.

A relatively strongly repulsive impurity which almost equally
couples to the two individual QD electrons will result in an entire
triplet state at B=0 due to the small competition between the two
e-I couplings. This can be observed by positioning the impurity
around the origin of the double-dot system. When the impurity is
found close to the bottoms of the confining potential well the
system will stay in the singlet state where the two electrons
perform the favored singlet state with spins anti-parallel to each
other. Due to this fact, the influence of the inter-dot separation
in the presence of impurity can be examined by modulating the well
barrier $V_b$ without changing the single-particle properties in
individual QDs. The reason is that the relative position of the
impurity with respect to the minima of the 2-dot system is
effectively controlled by changing $V_b$.  We found that an
attractive impurity serves to increase the ``tunnel" rate between
the two coupled dots.

In general, charged impurities ``destroy" the maximally entangled
singlet state by mixing the different singlets having varying
contributions to the total wave function. A strongly perturbative
impurity really ``messes up" the quantum information obtained from
this coupled QDs because the triplet state ``couples" to the second
excited state by the presence of anti-crossings with a slightly
larger than zero energy gap. The ground state now can favor a doubly
occupied state. Because the triplet is always obtained with
probability 1, it occurs in between the two lowest singlets: the
exchange singlet $\psi^S_1$ and the doubly-occupied $\psi^S_2$ or
$\psi^S_3$ depending on the sign of the effective charge $Z_1$. The
energy spectrum appears with anti-crossings as seen in our results.

In the presence of two identical charged impurities, the breakdown
between the two doubly-occupied singlets of the left and the right
dots is ``mended" if the impurities are located exactly
symmetrically along the line connecting the two confining potential
minima, i.e. the $x$-axis. As a result, one no longer obtains such a
bias between the two doubly-occupied singlets in the left- or the
right-dot as seen in the energy spectrum.

The mixing of different singlet states and anti-crossing coupling of
the singlet and triplet states due to the impurity presence results
in significant affect on the qubit operations, e.g. exchange and
C-NOT, in the considered coupled 2-quantum-dot system. %%

Finally, we discuss the implications of our results for the
operation of semiconductor spin qubits. Obviously, without some
quantitative knowledge of impurity locations near the qubits, no
real comparison between our results and experiments would be
possible.  But we can make some general remarks based on statistical
considerations.

First, semiconductor qubits are typically fabricated by lithographic
techniques creating QDs from parent 2DEG systems. The
low-temperature mobility of the 2DEG is controlled entirely by the
background charged impurity density both in Si\cite{Hwang} and in
GaAs\cite{Tracy} systems. Theoretical calculations\cite{Hwang,Tracy}
can provide quantitatively accurate information about the ensemble
averaged impurity density in the 2DEG from the mobility
measurements. This information about the background impurity density
can be converted to a statistical probability of finding impurities
located in specific QD structures since the effective active areas
of the dots would be known from the lithographic structures. For
example, typical GaAs 2DEG would have $10^{10}$  to $10^{11}$
charged impurities per sq. cm. whereas Si systems are typically
dirtier with $10^{11}$ to $10^{12}$ charged impurities per sq. cm.
This crudely translates to one charged impurity every 10 to 100 nm
in linear distance statistically. Typical lithographic qubits are
20-100 nm squares, indicating the presence of 1-5 charged impurities
per qubit statistically with GaAs (Si) being near the lower (higher
number) number. Of course, the details of impurity locations matter
very much, and the statistical considerations could be improved and
combined with our exchange splitting calculations to obtain the
effective sample ``yield" for a given 2DEG mobility, i.e. estimate
the fraction of samples that would statistically have the impurities
far enough for spin qubit operations to work.  But even these
elementary considerations show that the effective yield is likely to
be very low with probably only about 1-5\% of QD systems being
``lucky enough" to have the impurities located far enough from the
dots for them to work as spin qubits. The rest, even before taking
into account extraneous experimental factors (e.g. leakage, noise),
simply would not work because there is no effective singlet-triplet
coupling providing the necessary entanglement in the system. A
possible way of circumventing this problem may perhaps be having
several electrons per dot so that the external charged impurity
potential is effectively screened by the QD electrons, but this
raises the problem of having rather weak exchange coupling in
multielectron dots\cite{Hu}. The complex interplay of multielectron
dots in the presence of random charged impurities in the background
is left as an important future open problem in this subject, which
may very well be the next important step in this direction.

Second, an important effect of background charged impurities is
their fluctuations which produce various charge noise signals in
semiconductor devices. This charge noise may very well be the
limiting decoherence mechanism in currently existing semiconductor
QD spin qubits\cite{Yacoby}. Normally, of course, charge noise would
not adversely affect spin configurations, but semiconductor spin
qubits depend on the electrostatic exchange coupling which is
affected by external charged impurities as discussed extensively in
this paper. Therefore, any impurity fluctuations would lead to spin
qubit decoherence through the charge noise
mechanism\cite{Hu-charge,Culcer}. Our current work provides
quantitative estimate of the strength of such charge noise if the
fluctuation spectrum (or the fluctuation timescale) is known.
Deriving microscopic information about charge noise from our results
could be another future interesting direction of research.

Third, the inevitable presence of static charged impurities in the
background makes every semiconductor dot spin qubit, whether in GaAs
or in Si (or some other material), unique since the microscopic
electrostatic potential environment for each qubit will necessarily
differ in a random manner from qubit to qubit due to background
impurities as shown in this paper.  In particular, the
singlet-triplet energy separation and the consequent exchange
coupling will thus be somewhat random in a collection of many
qubits. This would in turn necessitate characterization of each
qubit in the eventual quantum computer rather precisely since the
gate operations depend very strongly on the knowledge of the
exchange coupling between the dots.  Such a characterization will
adversely affect the scalability of semiconductor spin quantum
computer architectures. Our work shows the importance of having
ultraclean impurity-free environment even for solid state quantum
computation, similar to the requirements for topological quantum
computation using non-Abelian fractional quantum Hall
states\cite{Nayak}, not because spin quantum computation needs
ultra-high mobilities\cite{Hwang}, but because it requires stable
values of exchange coupling without large qubit-to-qubit variations.
Our current work indicates that an order of magnitude reduction in
the background impurity concentration in GaAs, bringing it down to
around 10$^{12}$ per cubic cm, which is also the goal for
topological quantum computation\cite{Nayak}, should lead to the
production of 100-1000 spin qubits with impurity-free environment
rather easily. Unfortunately, for Si-SiO$_2$ systems this is a very
stringent condition because of the invariable presence of large
oxide charges near the interface\cite{Tracy}, but for
GaAs\cite{Hwang,Yacoby} and Si-Ge based quantum dot
systems,\cite{Qiuzi,Culcer,Eriksson} low-impurity materials may
become available for quantum computer architectures in the near
future.

Fourth, most of the spin qubit manipulation experiments in
semiconductor quantum dots are carried out under transport
situations using an external dc voltage bias. Such external voltage
affects the two dots differentially somewhat similar to the impurity
effects discussed in the current work since the confinement
potentials for the two dots are affected differently by the external
voltage. Our technique can be used to study this effect. Another
possible future direction of study could be impurity effects on the
full coupled quantum dot energy spectra (i.e. in addition to just
the double-dot exchange energy mainly considered in the current
work) since in some experiments\cite{Ono} higher energy levels come
into play. It will, in fact, be interesting also to consider
multielectron quantum dot systems with more than one electron per
dot and ask how impurity disorder affects the energy spectra. This
would necessitate a calculation of the quantum dot electronic
structure using many orbitals per dot (e.g. s, p, d, f, ...), which
is beyond the scope of the current work.

\section{Acknowledgments}
This work is supported by LPS-NSA and IARPA.

\appendix
\section{Coulomb matrix elements}\label{A:Coulomb}
\subsection{Single QDs} To evaluate the e-I
Coulomb matrix element, we calculate the integral:
\begin{equation}\label{e-Imp}
V^{n_2l_2}_{n_1l_1}\left(\widetilde{R}\right)=\int{\varphi^{*}_{n_1l_1}
\left(\bf{\widetilde{r}}\right)\frac{1}{|\bf{\widetilde{r}}-\bf{\widetilde{R}}|}
\varphi_{n_2l_2}\left(\bf{\widetilde{r}}\right) d\bf{\widetilde{r}}}
\end{equation}
where $\varphi_{n,l}({\widetilde{\textbf{r}}})$ is defined in Eq.
(\ref{e:Fock-Darwin}). The denominator in the right hand side of Eq.
(\ref{e-Imp}) is eliminated using the Gaussian identity:
\begin{equation}
\frac{1}{r}=\frac{2}{\pi}\int_{0}^{\infty}e^{-u^2r^2}du.
\end{equation}
Integrating integral (\ref{e-Imp}) over $\theta$ and set
$\widetilde{r}^2=t$, we arrive at:
\begin{eqnarray}\label{e-imp2}
V^{n_2l_2}_{n_1l_1}\left(\widetilde{R}\right)&=&\delta_{l_1,l_2}\sqrt{\frac{n_1!n_2!}{(n_1+l^{+})!(n_2+l^{+})!}}\\\nonumber
&&\int_{0}^{\infty}t^{l^{+}}e^{-(1+u^2)t}L_{n_1}^{l^{+}}\left(t\right)L_{n_2}^{l^{+}}\left(t\right)e^{-u^2\widetilde{R}^2}dtdu.
\end{eqnarray}
Next, we use one of the properties of the Laguerre
polynomials\cite{table}:
\begin{eqnarray}\nonumber
&&\int_{0}^{\infty}t^{\alpha-1}e^{-pt}L_{m}^{\lambda}\left({at}\right)L_{n}^{\beta}\left(bt\right)dt=\\\nonumber
&&\frac{\Gamma\left(\alpha\right)(\lambda+1)_m(\beta+1)_np^{-\alpha}}{m!n!}\sum_{j=0}^{m}\frac{(-m)_j(\alpha)_j}
{(\lambda+1)_jj!}{\left(\frac{a}{p}\right)}^{j}\\\nonumber
&&\sum_{k=0}^{n}\frac{(-n)_k(j+\alpha)_k}{(\beta+1)_kk!}\left(\frac{b}{p}\right)^k
\end{eqnarray}
and the gamma functions for integers to calculate the integral in
the right hand side of (\ref{e-imp2}). Finally, we arrive at the
formula presented in Eq.~(\ref{V_matrix})
\subsection{Coupled QDs}

In general, the Coulomb matrix elements between the electrons and
impurity are obtained numerically. Except for few cases with applied
constraint conditions to the impurity position, it is impossible to
obtain a closed algebraic form for this type of Coulomb interaction.
Using a similar method as before to calculate the e-I coupling
elements, i.e. $\langle\Psi|\widehat{V}_{\mbox{e-Imp}}|\Psi\rangle$,
we obtain:
\begin{eqnarray}\label{e:A_e-Iz}
\langle{\varphi_{L(R)}}|V_{\mbox{e-Imp}}|\varphi_{L(R)}\rangle=&\frac{ZV_0^C}{\sqrt{\pi}}
\int_{0}^{\infty}\frac{e^{-vz_0^2-\frac{v}{v+1}}}{\sqrt{v}(v+1)}dv\\\nonumber
\langle{\varphi_{L(R)}}|V_{\mbox{e-Imp}}|\varphi_{R(L)}\rangle=&ZV_0^C\sqrt{\pi}
e^{z_0^2-a_0^2} \mbox{erfc} (z_0)
\end{eqnarray}
for $\textbf{R}=(0,0,z)$ and
\begin{eqnarray}\label{e:A_Ixy}
\langle{\varphi_{L(R)}}|V_{\mbox{e-Imp}}|\varphi_{L(R)}\rangle=&ZV_0^C\sqrt{\pi}
e^{-\frac{[(x_0\pm a_0)^2+y_0^2]}{2}}\\\nonumber &I_0[\frac{(x_0\pm
a_0)^2+y_0^2}{2}]\\\nonumber
\langle{\varphi_{L(R)}}|V_{\mbox{e-Imp}}|\varphi_{R(L)}\rangle
=&ZV_0^C\sqrt{\pi} e^{-[a_0^2-\frac{(x_0^2+y_0^20}{2}]}\\\nonumber
&I_0[\frac{(x_0^2+y_0^2)}{2}]
\end{eqnarray}
for $\textbf{R}=(x_0,y_0,0)$ where $a_0=a_m/l_0$, $x_{0}=x/l_0,
y_0=y/l_0,$ and $z_0=z/l_0$.

It is of interest to consider the special situation when the
impurity is located along the line connecting the two potential well
minima, i.e. along the $x$-axis in this case. In the above sections,
we discussed this case in depth. The Coulomb matrix elements for the
left and the right QD electron are as follows:
\begin{eqnarray}
\langle{\varphi_{L(R)}}|V_{\mbox{e-Imp}}|\varphi_{L(R)}\rangle&=&ZV_0^C\sqrt{\pi}
e^{-\frac{(x_0\pm a_0)^2}{2}}\nonumber\\&&I_0[\frac{(x_0\pm
a_0)^2}{2}]\\\nonumber
\langle{\varphi_{L(R)}}|V_{\mbox{e-Imp}}|\varphi_{R(L)}\rangle&=&ZV_0^C\sqrt{\pi}
e^{-(a_0^2+\frac{x_0^2}{2})}I_0[\frac{x_0^2}{2}].
\end{eqnarray}
%As the impurity effective charge is increased, the coupling between
%electron and the impurity becomes dominant. The triplet and
%doubly-occupied singlets become virtually ``couple" with each other
%via a slightly larger than zero energy gap when the impurity is
%found around the well bottoms. The ground state turns out to be a
%competing situation between the spin-exchange singlet $\Psi_1^S$ and
%either of the two other singlets $\{\psi^S_2,\psi^S_3\}$. As a
%result, the singlet-triplet splitting $J$ exhibits a maximum at the
%well bottoms. When higher Fock-Darwin states are included, this
%maximal behavior is still found at the well bottoms. This is opposed
%to the small $Z_1$, e.g. $Z_1=\mp0.1$, case where such extrema in
%$J$ change when higher excited Fock-Darwin states are included.

\section{Heitler-London approximation}\label{A:HL}

The closed analytical form of the exchange energy between the two
coupled QDs for a single charged impurity located arbitrarily in the
$xy$-plane [$\textbf{R}=(x_0,y_0,0)l_0$] is obtained by solving a
basic two-level problem:
\begin{widetext}
\begin{eqnarray}\label{e:JHL}
J_{\mbox{HL}}&=&-\frac{e^{a_0^2}}{1-e^{4a_0^2}} \Big{\{} 2
\hbar\omega_0 a_0^2e^{a_0^2}+ \frac{4 V_0 (e^{\frac{a_0^2 - 2 a_0
a_{0x}^2 a_8 - a_{0x}^2 a_8^2}{1 + a_{0x}^2}} + e^{\frac{a_0^2 + 2
a_0 a_{0x}^2 a_8 - a_{0x}^2 a_8^2}{1 + a_{0x}^2}} - 2
e^{a_0^2-\frac{a_{0x}^2a_8^2}{1+a_{0x}^2}}
)}{\sqrt{(1+a_{0x}^2)(1+a_{0y}^2)}}\\\nonumber &&
\,\,\,\,\,\,\,\,\,\,\,\,\,\,\,\,\,\,\,\,\,\,\,\,\, + \frac{4 V_b
(e^{\frac{a_0^2}{1+a_{0bx}^2}}-e^{a_0^2})}{\sqrt{(1 + a_{0bx}^2)(1 +
a_{0by}^2)}} + \sqrt{\pi} V_0^C \Big{[}- \sqrt{2}e^{a_0^2} +
\sqrt{2} I_0(a_0^2) - 4 Z e^{a_0^2 - \frac{x_{0}^2 + y_{0}^2}{2}}
I_0(\frac{x_{0}^2+y_{0}^2}{2})
\\\nonumber && \,\,\,\,\,\,\,\,\,\,\,\,\,\,\,\,\,\,\,\,\,\,\,\,\, +
2 e^{\frac{a_0^2 + 2 a_0 x_{0} - x_{0}^2 - y_{0}^2}{2}} Z
I_0(\frac{a_0^2 - 2 a_0 x_{0} + x_{0}^2 + y_{0}^2}{2})  + 2
e^{\frac{a_0^2 - 2 a_0 x_{0} - x_{0}^2 - y_{0}^2}{2}} Z
I_0(\frac{a_0^2 + 2 a_0 x_{0} + x_{0}^2 + y_{0}^2}{2}) \Big{]}
\Big{\}}.
\end{eqnarray}
\end{widetext}
Here, $a_8=a/l_0$, $a_{0x}=l_0/l_x$, $a_{0y}=l_0/l_y$,
$a_{0bx}=l_0/l_{bx}$, and $a_{0by}=l_0/l_{by}$ are dimensionless
parameters associated with a series of lengths such as $a, l_x, l_y,
l_{bx}, l_{by}$. These parameters together with depths $V_0$ and
$V_b$ define the size and the shape of the confinement double well.
\begin{figure}[hbtp]
\begin{center}
\vspace*{-0.5cm}\leftskip0.5cm
\includegraphics[width=7.2cm]{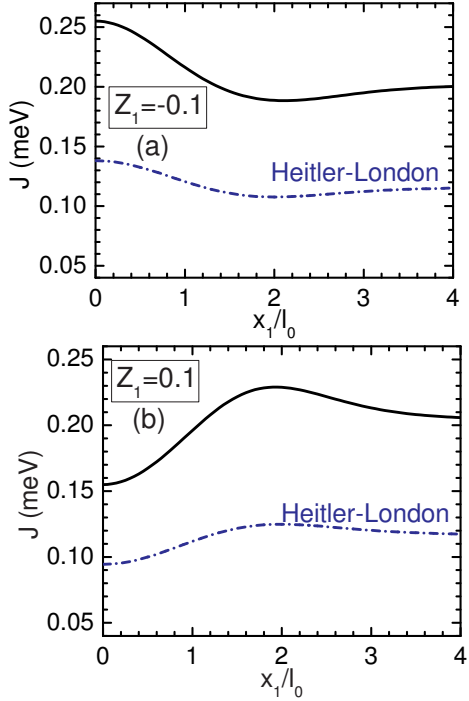}
\end{center}
\vspace{-0.8cm} \caption{(Color online) Singlet-triplet spin
splitting $J$ as a function of the impurity position along the
$x$-axis for (a) $Z_1$=-0.1 and (b) 0.1. Black solid curves are
extracted from Fig.~\ref{double-dot_Z1M1_x1_B=0_ST} and blue dashed
curves are obtained from Eq.~(\ref{e:JHL}). We receive qualitative
agreement between our numerical calculations and analytical results
where the singlet-triplet splitting energy calculated using these
two methods exhibits a (a) minimum for $Z_1$=-0.1 and (b) maximum
for $Z_1=0.1$ around $x_1\approx2l_0$.}
 \label{double-dot_Vb30_Z1=-0.1_B=0_HL}
\end{figure}
\begin{figure}[hbtp]
\begin{center}
\vspace*{-0.5cm}\leftskip-0.0cm
\includegraphics[width=7.5cm]{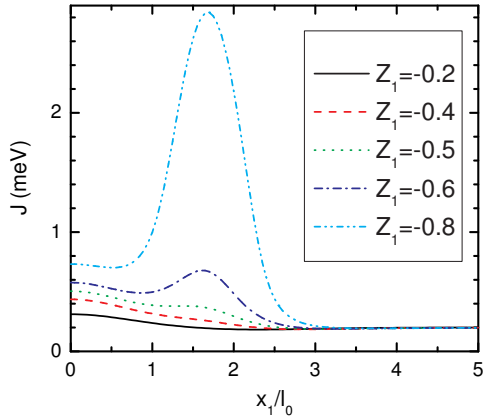}
\end{center}
\vspace{-0.8cm} \caption{(Color online) Singlet-triplet spin
splitting $J$ as a function of the impurity position along the
$x$-axis for different $Z_1<0$. In this plot one can see the gradual
change in $J$ from having a minimum around $x_1\approx2l_0$ to
having a maximum at different $x_1$ depending on $Z_1$ as $Z_1$ is
increases from -0.1 (black solid) to -0.8 (cyan dash-dotted-dot).}
\label{double-dot_Vb30_Z1=2468_J}
\end{figure}
%%
%%%%%
%%
\begin{figure}[hbtp]
\begin{center}
\vspace*{-0.5cm}\leftskip-0.0cm
\includegraphics[width=7.5cm]{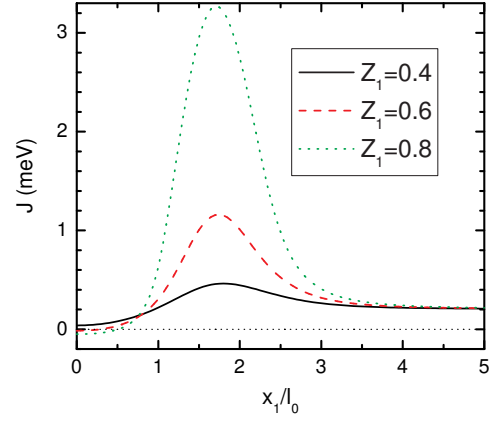}
\end{center}
\vspace{-1.0cm} \caption{(Color online) Singlet-triplet spin
splitting $J$ as a function of the impurity position along the
$x$-axis for different $Z_1>0$. The maximum in $J$ is found at
different $x_1$ as $Z_1$ changes. } \label{double-dot_Vb30_Z1=468_J}
\end{figure}

Heitler-London approximation does not allow one to examine the
impurity effect on the coupled qubit operations manifested in the
mixing of the different singlets and ``coupling" between the
singlet-triplet states as already observed and discussed in e.g.
Fig.~\ref{double-dot_Z1MP1_Vb30_x_trum}(b) The impurity-position
dependence of $J_{HL}$, as obtained analytically in the last three
terms in Eq.~(\ref{e:JHL}), now can be studied in
Fig.~\ref{double-dot_Vb30_Z1=-0.1_B=0_HL} as a function of $x_0$. We
see that the presence of the charged impurity only plays a role as a
weak perturbative interaction to the total energy even in case the
effective e-I coupling is found relatively large (the case
$Z_1=-1$). Such a property like the triplet-singlet transition for
weak repulsive e-I exchange interaction is not found in this case.

\section{Singlet-triplet splitting $J$ for intermediate impurity effective
charges}\label{A:Intermediate}

\begin{figure}[hbtp]
\begin{center}
\vspace*{-0.1cm}\leftskip-0.0cm
\includegraphics[width=7.5cm]{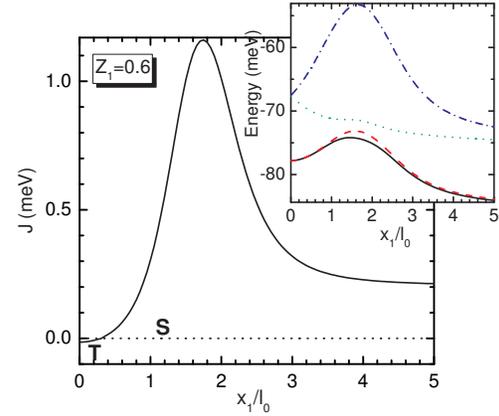}
\end{center}
\vspace{-0.8cm} \caption{(Color online) Singlet-triplet spin
splitting $J$ as a function of the impurity position along the
$x$-axis for a specific case of $Z_1>0$: 0.6. Inset is the low-level
energy spectrum of the system. As compared to the pronounced
anti-crossing seen in Fig.~\ref{double-dot_Z1MP1_Vb30_x_trum}(b),
the $Z_1=0.6$ case has the spectrum which is close to having
pronounced anti-crossings as obtained for $Z_1=1$ in
Fig.~\ref{double-dot_Z1MP1_Vb30_x_trum}(b).}
\label{double-dot_Vb30_Z1=0.6b_J}
\end{figure}

We show the results obtained for several different intermediate
$Z_1$ for both the negatively and positively charged impurity cases
in Figs.~\ref{double-dot_Vb30_Z1=2468_J},
~\ref{double-dot_Vb30_Z1=468_J}, and
~\ref{double-dot_Vb30_Z1=0.6b_J}. %%
These plots provide a clearer and deeper look into the
triplet-singlet transition (Figs.~\ref{double-dot_Vb30_Z1=468_J} and
~\ref{double-dot_Vb30_Z1=0.6b_J}) for a strongly repulsive impurity
as well as the physics of anti-crossings found in the energy
spectrum of the coupled 2-dot system. For the $Z_1<0$ case,
Fig.~\ref{double-dot_Vb30_Z1=2468_J} serves to explain in detail the
transition from having a minimum to a maximum in the singlet-triplet
splitting $J$ between the two lowest energy levels.

\section{Spin splitting energy of Si/SiGe coupled quantum
dots}\label{A:Si}

Si QDs have a relatively large effective mass $m_{Si}^{*}\approx2.8
m_{GaAs}^*$. Plus, the effective Rydberg energy in Si QDs, as
discussed in Sec. II, is higher than the effective Rydberg energy of
GaAs QDs. These bring into the fact that the Si QDs have a smaller
electron kinetic energy, which means that the electrons tend to be
more localized as compared to the case of GaAs. Thus the tunnel rate
is expected to be smaller. Our calculations show that for the same
model of a coupled 2-dot system in which the shape and the size and
the barrier height are kept unchanged: e.g. $V_0=-50$ meV and
$V_b=30$ meV (corresponding to $\Delta{V_b}=9.65$ meV), Si/SiGe
double-dot system has $\hbar\omega_0\approx6.67$ meV
($l_0\approx8.48$ nm). The spin singlet-triplet splitting $J$
between the two lowest energy levels in Si/SiGe double dots is found
$J\approx0.003$ meV, much smaller than the $J\approx0.204$ meV of
the GaAs/AlGaAs double dots. In experiments, the electron tunneling
rate in Si/SiGe coupled double QDs is indeed found much lower than
the tunneling rate in GaAs/AlGaAs coupled QDs (see e.g.
Ref.\cite{Eriksson}). In the presence of a charged impurity, e.g.
$Z_1=-0.1$ located at the origin, we obtained $J$ increased to
$J\approx0.008$ meV for Si/SiGe double QDs as compared to
$J\approx0.26$ meV for GaAs/AlGaAs double QDs.

%\begin{references}


\begin{thebibliography}{99}
\bibitem{note1} The minimum in $J$ varies when $V_b$ changes.
Therefore, the minimal splitting energy for $V_b=13$ meV is found at
a different position of the impurity as compared to the ``standard"
$\approx 2l_0$ for the case $V_b=30$ meV
\end{thebibliography}
\end{document}